\documentclass[11pt]{article}
\usepackage[final]{acl}

\usepackage[T1]{fontenc}
\usepackage[utf8]{inputenc}
\usepackage{times}
\usepackage{latexsym}
\usepackage{microtype}
\usepackage{graphicx}
\usepackage{wrapfig}
\usepackage{subcaption}
\usepackage{booktabs} % for professional tables

\usepackage{algorithm}
\usepackage{algorithmic}

\usepackage{amsmath}
\usepackage{amssymb}
\usepackage{mathtools}
\usepackage{amsthm}

\usepackage{enumitem}
\usepackage{listings}
\usepackage{multirow}
\usepackage{makecell}
\usepackage{xcolor} % Already loaded by icml2026, but good to be aware

\usepackage{listings}
\usepackage{xcolor}

\definecolor{codebg}{HTML}{FAFAFA}
\definecolor{codeframe}{HTML}{D8D8D8}
\definecolor{codekw}{HTML}{0F5C9E}   % keywords
\definecolor{codestr}{HTML}{2E7D32}  % strings
\definecolor{codecom}{HTML}{8A8A8A}  % comments
\definecolor{codenum}{HTML}{B0B0B0}  % line numbers
\definecolor{codefn}{HTML}{7B3FA0}   % function names / builtins

\lstdefinestyle{pythonpretty}{
  language=Python,
  basicstyle=\ttfamily\tiny,
  keywordstyle=\color{codekw}\bfseries,
  keywordstyle=[2]\color{codefn},
  stringstyle=\color{codestr},
  commentstyle=\color{codecom}\itshape,
  numbers=left,
  numberstyle=\ttfamily\tiny\color{codenum},
  numbersep=8pt,
  backgroundcolor=\color{codebg},
  frame=single,
  rulecolor=\color{codeframe},
  framerule=0.4pt,
  framexleftmargin=4pt,
  framexrightmargin=0pt,
  xleftmargin=14pt,
  aboveskip=1em,
  belowskip=1em,
  breaklines=true,
  breakatwhitespace=true,
  breakindent=1.5em,
  postbreak=\mbox{\textcolor{codenum}{$\hookrightarrow$}\space},
  showstringspaces=false,
  columns=fullflexible,
  keepspaces=true,
  tabsize=4,
  upquote=true,
  morekeywords=[2]{self, format, print, None, True, False},
}

\lstdefinestyle{cpretty}{
  language=C,
  basicstyle=\ttfamily\tiny,
  keywordstyle=\color{codekw}\bfseries,
  keywordstyle=[2]\color{codefn},
  stringstyle=\color{codestr},
  commentstyle=\color{codecom}\itshape,
  numbers=left,
  numberstyle=\ttfamily\tiny\color{codenum},
  numbersep=8pt,
  backgroundcolor=\color{codebg},
  frame=single,
  rulecolor=\color{codeframe},
  framerule=0.4pt,
  framexleftmargin=4pt,
  framexrightmargin=0pt,
  xleftmargin=14pt,
  aboveskip=1em,
  belowskip=1em,
  breaklines=true,
  breakatwhitespace=true,
  breakindent=1.5em,
  postbreak=\mbox{\textcolor{codenum}{$\hookrightarrow$}\space},
  showstringspaces=false,
  columns=fullflexible,
  keepspaces=true,
  tabsize=4,
  upquote=true,
  morekeywords=[2]{self, format, print, None, True, False},
}

\lstdefinestyle{moderncstyle}{
    language=C,
    basicstyle=\ttfamily\small,
    keywordstyle=\color{blue}\bfseries,
    commentstyle=\color{green!60!black},
    stringstyle=\color{red},
    showstringspaces=false,
    breaklines=true,
    frame=single
}

\definecolor{codegray}{rgb}{0.5,0.5,0.5}
\definecolor{codeblue}{rgb}{0.13,0.13,1}
\definecolor{codestring}{rgb}{0.63,0.125,0.094}
\usepackage{soul}
\usepackage{siunitx}

\usepackage[table]{xcolor}

\definecolor{tablerowhighlight}{gray}{0.9}

\usepackage{amsmath}
\sethlcolor{tablerowhighlight}
\usepackage[dvipsnames]{xcolor}
\usepackage{tabularx} % REQUIRED: Add this to your preamble

\usepackage{soul}

\usepackage[capitalize,noabbrev]{cleveref}

\theoremstyle{plain}

\theoremstyle{definition}

\theoremstyle{remark}

\usepackage[textsize=tiny]{todonotes}

\title{Vulnerable Code Search: Transferable Attack for Code Language Models}

\begin{document}

\author{%
  Kaicheng Wang \quad Liyan Huang \quad Jesse Thomason \quad Weihang Wang \\
  University of Southern California \\
  \texttt{\{wangkaic, liyanhua, jessetho, weihangw\}@usc.edu}}

\maketitle

\begin{abstract}
Reliable code retrieval is crucial for developer productivity and effective code reuse. However, current neural code language models (CLMs) powering search tools are susceptible to adversarial attacks targeting non-functional textual elements. In this paper, we introduce a programming language-agnostic, transferable, adversarial attack that exploits this CLM vulnerability. Our approach perturbs identifiers within a code snippet without altering the snippet's functionality to artificially align the code with a target query. We demonstrate that our attack, even when computed using smaller code embedding models, such as CodeT5+, is highly effective and transferable to larger, closed-source embedding models, like Voyage-code-3, or LLMs like Gemini-3.1-Pro. Our attack can increase the similarity between the query and arbitrary, irrelevant code snippets, consequently degrading key retrieval metrics such as the Mean Reciprocal Rank (MRR) of state-of-the-art models by up to 77\%. The experimental results highlight the fragility of current code search methods and underscore the need for more robust, semantic-aware approaches.

% Our codebase is available at \url{https://github.com/AdvAttackOnNCC/Code_Search_Adversarial_Attack}. \kw{Remember to remove the codebase links.}
\end{abstract}

\section{Introduction}

% \kw{@Dr. Wang, I planned to add a table comparing our work with the previous work here, but changed my mind for 2 reasons. 1) We only compare with 2 baselines in the later section. Mentioning a lot other works here lead to questions from reviewers; 2) the tight page limit.}

\begin{figure}[t]
    \centering % Center the entire figure content
    % \Description{Overview diagram of the proposed adversarial attack pipeline, showing how a natural-language query and a code snippet are processed and how identifier substitutions increase embedding similarity without changing functionality.}
    \includegraphics[width=1.0\linewidth]{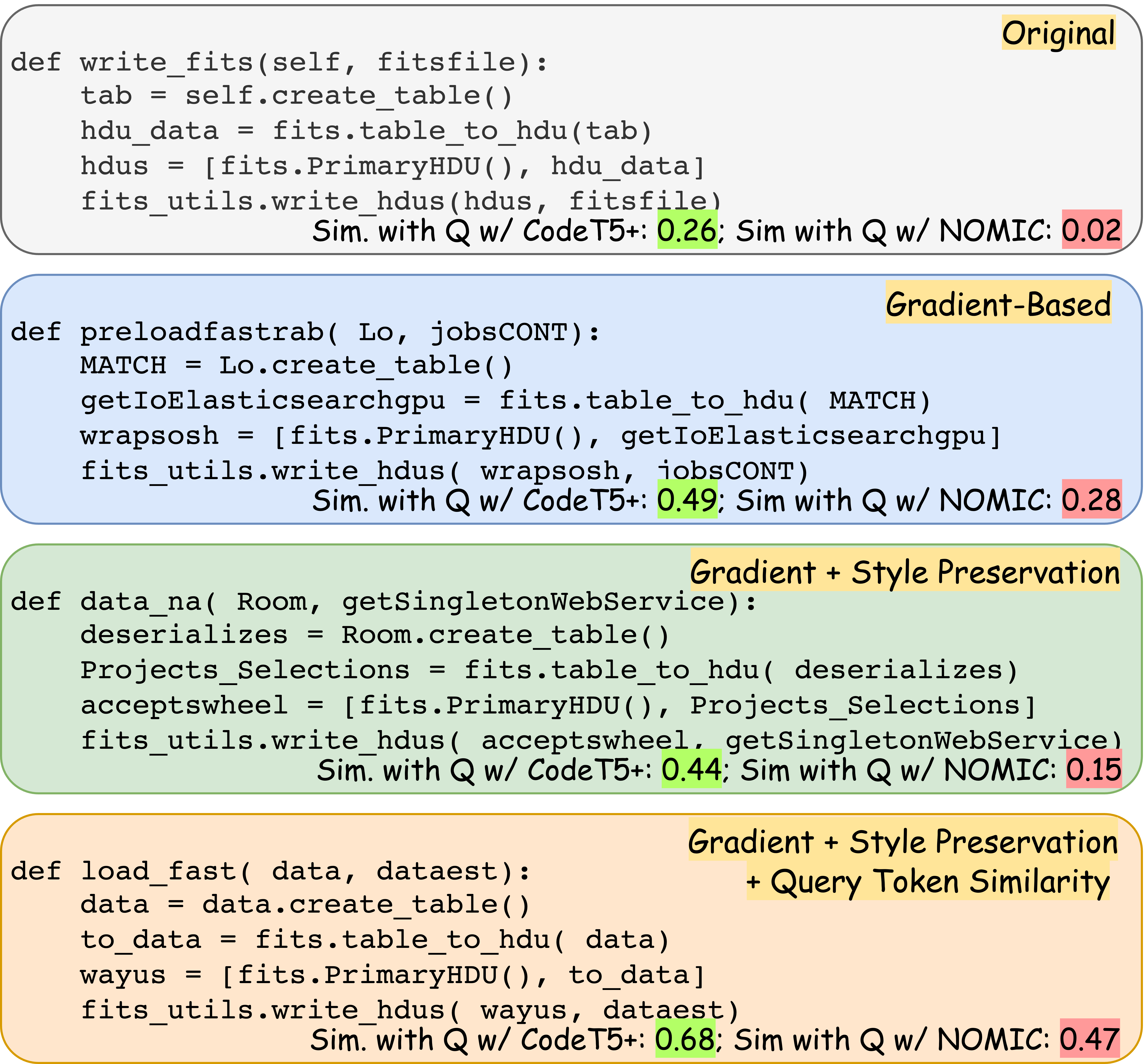}
\caption{Overview of our attack for code search. Given the target query ``\texttt{fastest way to load data}'', the attack replaces identifiers in an unrelated code snippet while preserving functionality to deceive code search models. All attack variants are optimized on CodeT5+, but the resulting adversarial code snippets transfer beyond the surrogate model, also increasing similarity on the larger Nomic-embed-code model.}
\label{fig:method_illustration} 
\end{figure}

The rapid expansion of the software ecosystem has intensified the need for efficient code retrieval to improve developers' productivity~\cite{ai_for_swe, code_search_survey_2023, coir}. While the advent of Large Language Models (LLMs) and specialized Code Language Models (CLMs) has revolutionized tasks like repository-level reasoning~\citep{CLM_survey, CodeLlama, qwen2.5, Codex, repoqa, swebench, repobench}, their computational cost often prohibits direct application in large-scale retrieval scenarios involving millions of candidates~\citep{google_loc, llm_ecnomic}. Consequently, for practical and efficient search, the industry continues relying on embedding-based methodologies, which encode queries and code into a shared latent space to rank candidates based on vector similarity~\citep{code_search_survey_michael_pradel, challenge_for_code_search}.

However, embedding-based retrieval is susceptible to a critical vulnerability: modern CLMs are vulnerable to adversarial examples---code snippets containing small, functionality-preserving modifications that can drastically change the resulting embeddings~\citep{AdvAttack_survey, advattack_survey2, adv_attack_backdoor_survey_2024}. While most prior research has focused on adversarial attacks in classification tasks~\citep{DAMP, ACCENT, carl, dip}, our work centers on code search, which leads to unique challenges. First, adversarial attacks in code search can be tailored to specific queries. By targeting specific queries, adversaries can have irrelevant or malicious code snippets crowding out legitimate results during certain task inquiries, subsequently influencing downstream coding decisions. Second, the typical code search workflow introduces additional robustness issues. Unlike classification, where malicious code is processed directly during inference and can be inspected by LLMs with reasoning abilities~\citep{defect_detection_on_advattack_on_code, llm_detect_malicious_code, llm_swe_security_survey}, code search largely relies on offline embedding. When large code corpora are embedded offline, the modified code snippets maintain valid semantics and appear harmless. Since the subsequent retrieval process operates purely based on the similarity of these precomputed embeddings, detecting or mitigating these adversarial inputs becomes significantly more difficult.

In this paper, we demonstrate the shared vulnerability of the CLMs through transferable adversarial attacks. We generate adversarial examples by strategically replacing identifiers---user-defined symbols including function and variable names---within a code snippet to maximize the snippet's embedding similarity to a target query. The replacements are usually optimized on small white-box CLMs using gradient information, identifier style preservation, and query token similarity. As illustrated in Figure~\ref{fig:method_illustration}, although the attacks are crafted on \texttt{CodeT5+}, the resulting similarity improvement also appears under \texttt{Nomic-embed-code}, a model roughly 60 times larger. Based on our experiments, the attack transferability is broader: adversarial snippets generated on small surrogate models can transfer to larger models (e.g., \texttt{OASIS}, \texttt{Nomic-embed-code}) and closed-source systems, including embedding models (e.g., \texttt{Voyage-code-3}) and LLMs (e.g., \texttt{Gemini-3.1-Pro}) across five programming languages. Furthermore, the similarity change on the small source model strongly correlates with the changes on larger or closed-source target models, suggesting that attack effectiveness can be efficiently estimated.

Although state-of-the-art models achieve high scores on standard code search benchmarks~\citep{COIR_leaderboard, benchmark_saturate}, they are vulnerable to our transferable adversarial attacks, leading to catastrophic drops in key retrieval metrics. For instance, we observed an absolute drop of up to 77\% in Mean Reciprocal Rank (MRR) across all tested models on CosQA~\cite{cosqa}.
The performance degradation exposes a critical gap: high benchmark scores do not necessarily reflect true semantic understanding of code. Instead, they highlight a reliance on brittle lexical features, underscoring the urgent need for improved model robustness.

\noindent In summary, our contributions are as follows:
\begin{itemize}[nosep,noitemsep,leftmargin=*]
% \begin{itemize}[nosep]
    \item We propose an adversarial attack method for code search, which perturbs the code snippet to maximize its similarity with the target query while preserving functionality;
    \item We demonstrate the strong transferability of the attack: adversarial code snippets generated using small models can effectively deceive larger or closed-source systems; and
    \item We reveal that high benchmark scores mask a critical lack of robustness: current models prioritize superficial lexical matching over semantic reasoning, 
    resulting in severe vulnerabilities that standard evaluations fail to capture.
\end{itemize}

\section{Related Work}

The field of Code Language Models (CLMs) has evolved rapidly, from early unified representations like CodeBERT~\citep{CodeBERT} to encoder-decoders such as CodeT5 and CodeT5+~\citep{CodeT5, codet5+}. Subsequently, large generative CLMs such as Codex~\citep{Codex} and numerous open-source efforts, including CodeLlama~\citep{CodeLlama} and StarCoder~\citep{starcoder, starcodev2}, have demonstrated sophisticated code generation and reasoning abilities. The generative capabilities of CLMs have been further enhanced by Retrieval-Augmented Generation (RAG)~\citep{general_rag_example}. In RAG systems, CLMs leverage efficiently retrieved code snippets---often sourced via embedding models---to improve contextual relevance and accuracy for complex tasks~\citep{coderag,code_search_for_rag_example1, code_search_for_rag_example2, code_search_for_rag_example3}. The robustness of these underlying code embedding models is therefore critical, as the vulnerabilities explored in this paper directly threaten downstream applications.

Code search, a crucial task for efficient software development and code reuse~\citep{CodeSearchSWE}, relies on embedding models for effective retrieval from large codebases. CodeSearchNet~\citep{CodeSearchNet} established early benchmarks, with models like CodeBERT~\citep{CodeBERT} learning joint natural language-code representations. To enhance understanding, later work incorporated structural information: GraphCodeBERT~\citep{graphcodebert} used data flow graphs, while UniXcoder~\citep{unixcoder} and SynCoBERT~\citep{syncobert} leveraged Abstract Syntax Trees. Later, contrastive learning, as seen in ContraCode~\citep{ContraCode}, became a dominant training technique. Other strategies include adapting general CLM embeddings~\citep{codet5+}, fine-tuning LLMs~\citep{nomic-embed-code}, or training on augmented data~\citep{oasis}. Recently, black-box embedding services provided by OpenAI~\citep{openai_text_embedding} and Voyage AI~\citep{voyage_code_3} have also achieved state-of-the-art performance. However, our experimental results demonstrate that high benchmark scores do not guarantee robustness; top-performing code search models remain susceptible to the proposed adversarial attacks.

Applying gradient-based attack methods~\citep{CVAdversarial} directly to natural language is challenging due to the discrete nature of tokens, making it difficult to apply small perturbations while maintaining syntactic and semantic integrity~\citep{adv_attack_nlp_general_survey}. Programming languages, however, provide unique opportunities for adversarial attacks through semantic-preserving transformations~\citep{defect_detection_on_advattack_on_code}. Attack strategies vary based on the assumed knowledge of the target model. White-box attacks require access to model gradients. Methods like DAMP~\citep{DAMP}, MHM~\citep{mhm}, and GraphCodeAttack~\citep{graph_code_attack} leverage gradient signals to guide modifications aimed at causing misclassification. Without access to internal model information, black-box attacks employ various optimization techniques. For instance, CARL~\citep{carl} and ALERT~\citep{ALERT} apply reinforcement learning and genetic algorithms, respectively. Meanwhile, CodeAttack~\cite{codeattack} relies on repetitive querying to pinpoint vulnerable tokens, and \citet{variable_renaming_adv_test_gen} investigates the efficacy of combined search strategies. Other black-box methods rely on heuristics, such as inserting comments or dead code~\citep{dip}. Another emerging approach uses generative models to directly produce adversarial code examples, as explored by CBA~\citep{cba} and ITGen~\citep{itreative_adv_attack}. In this work, we propose a novel adversarial attack method that leverages white-box attack techniques on one code search model to derive examples that also serve as transferable black-box attacks on other models.

\newcommand{\Qemb}{Q_{\text{emb}}}
\newcommand{\Cemb}{C_{\text{emb}}}
\newcommand{\Sim}{\text{Sim}}
\newcommand{\AAttack}{\textsc{AdvAttack}}
\newcommand{\LCS}{\text{LCS}}
\newcommand{\Remove}{\text{Remove}}

\section{Adversarial Attack for Code Search}
\label{sec: adversarial_attack_for_code_search}

% \begin{figure*}[t]
%     \centering % Center the entire figure content
%     % \Description{Overview diagram of the proposed adversarial attack pipeline, showing how a natural-language query and a code snippet are processed and how identifier substitutions increase embedding similarity without changing functionality.}
%     \includegraphics[width=0.9\linewidth]{plots/adv_attack_illu.drawio.png}
% \caption{Illustration of the adversarial attack methods detailed in Sections~\ref{ssec: gradient_based_method}, \ref{ssec: preserving_identifier_format}, and \ref{ssec: incorporating_query_similarity}. The target query is ``\texttt{fastest way to load data}'', and similarity scores are calculated using CodeT5+. The figure demonstrates that combining style preservation and query token similarity with the gradient approach leads to the most effective attack (highest similarity) while making the perturbations more natural and harder to notice.}
%     \label{fig:method_illustration} 
% \end{figure*}

In this section, we introduce our adversarial attack method against code search models. We first outline the threat model underlying our approach (Section~\ref{ssec: threat_model}). Subsequently, we detail the proposed methodology, including the use of gradient information to search for replacements (Section~\ref{ssec: gradient_based_method}), the strategy for preserving identifier style (Section~\ref{ssec: preserving_identifier_format}), and the incorporation of query token similarity (Section~\ref{ssec: incorporating_query_similarity}). Finally, we discuss the application of our method in black-box scenarios via attack transfer (Section~\ref{ssec: attack_transfer_method}).

\subsection{Threat Model}
\label{ssec: threat_model}

\textbf{Attack Goals}: Given a natural language query $Q$ and a target code snippet $C$ (e.g., malicious or semantically unrelated to $Q$), the adversary aims to manipulate the retrieval ranking by generating an adversarial variant $C'$ optimized to rank within the top-$k$ retrieval results for $Q$.

\textbf{Attacker Knowledge}: The adversary has full white-box access to a code embedding model, referred to as the \textit{surrogate model}. The surrogate model may differ from the actual victim model used by the code search system. The attacker can use the surrogate model to compute the similarity between $Q$ and $C$, along with the gradient of the similarity score with respect to the input token embeddings, to guide the generation of the adversarial example.

\textbf{Attacker Capability \& Practicality}: The attack targets the indexing phase, requiring the adversary to inject or modify code in the retrieval corpus. This is realistic in open-source settings, where corpus poisoning and software supply-chain attacks can occur through fraudulent repositories, malicious commits, or compromised maintainers~\cite{adv_attack_backdoor_survey_2024}.
%, practicality_1, practicality_2, practicality_4}. 
Once such code is indexed, the adversarial variant $C'$ can influence future retrieval results. 
%\kw{refer to the appendix for details}

\textbf{Attack Constraints}: To ensure the attack remains functional, code modification must not alter the execution logic of the target code $C$. Semantic preservation of the code snippet is ensured by maintaining the Abstract Syntax Tree (AST) structure unchanged before and after the attack.

\subsection{Gradient-Guided Optimization}
\label{ssec: gradient_based_method}

\textbf{Gradient-Based Approximation.}
Let $\Qemb$ and $\Cemb$ denote the initial embeddings for query $Q$ and code snippet $C$ after tokenization and mapping. We define the similarity score as $\Sim_\theta(Q, C) = G_\theta(\Qemb, \Cemb)$, where $\theta$ denotes the surrogate model parameters. Our goal is to find a modified snippet $C'$ that maximizes the similarity $\Sim_\theta(Q, C')$.

Let $\delta = C'_{\text{emb}} - \Cemb$. By applying a first-order Taylor expansion around $\Cemb$, the similarity change can be approximated as:
% \begin{align}
% \Delta \Sim_\theta &= G_\theta(\Qemb, C'_{\text{emb}}) - G_\theta(\Qemb, \Cemb) \nonumber \\
% &\approx \delta^{\top} \nabla_{\Cemb} G_\theta(\Qemb, \Cemb).
% \label{eq:taylor_approx}
% \end{align}
\begin{equation}
\begin{split}
\Delta \Sim_\theta
&= G_\theta(\Qemb, C'_{\text{emb}}) - G_\theta(\Qemb, \Cemb) \\
&\approx \delta^{\top} \nabla_{\Cemb} G_\theta(\Qemb, \Cemb).
\end{split}
\label{eq:taylor_approx}
\end{equation}

Here, the zeroth-order terms $G_\theta(\Qemb, \Cemb)$ cancel out, leaving the gradient term to approximate the change. The gradient $\nabla_{\Cemb} G_\theta(\Qemb, \Cemb)$ is a matrix of shape $c \times d$, where $c$ is the context length and $d$ is the embedding dimension.

\textbf{Greedy Search.}
Consider an identifier $w$ in the code snippet $C$. Let $t$ be a token in $w$, and let $\mathcal{P}_{t} = \{p_1, \dots, p_k\}$ denote the set of indices where $t$ appears as part of the identifier $w$ in the input token sequence. Note that $\mathcal{P}_{t}$ excludes positions where $t$ appears as a standalone keyword or within other identifiers.

When substituting $t$ with a candidate token $t' \in V$, the perturbation vector $\delta$ is sparse. It is non-zero only at indices $i \in \mathcal{P}_{t}$, where each entry corresponds to the difference between the candidate and original token embeddings $\mathbf{e}_{t'}$ and $\mathbf{e}_{t}$:
\begin{equation*}
    \delta[i] = 
\begin{cases} 
   \mathbf{e}_{t'} - \mathbf{e}_{t} & \text{if } i \in \mathcal{P}_{t}, \\
   0 & \text{otherwise.}
\end{cases}
\end{equation*}

By substituting the sparse $\delta$ into Eq.~\ref{eq:taylor_approx}, we isolate the contribution of this token replacement. Since the gradient at index $i$, denoted $\nabla_{\Cemb} G_\theta(\Qemb, \Cemb)[i]$, is a vector of shape $d \times 1$, we can compute the scalar contribution via a dot product with the $1 \times d$ perturbation vector $(\mathbf{e}_{t'} - \mathbf{e}_{t})^{\top}$. We define the total contribution as the \textit{influence}:
% \begin{equation}
% \label{eq:influence_def}
% \begin{split}
% &\textit{influence}(t, t', \mathcal{P}_{t}) \\
% &\quad = (\mathbf{e}_{t'} - \mathbf{e}_{t})^{\top} \left( \sum_{i \in \mathcal{P}_{t}} \nabla_{\Cemb} G_\theta(\Qemb, \Cemb)[i] \right),
% \end{split}
% \end{equation}
\begin{equation}
\label{eq:influence_def}
\begin{split}
&\textit{influence}(t, t', \mathcal{P}_{t}) \\&= (\mathbf{e}_{t'} - \mathbf{e}_{t})^{\top} \left( \sum_{i \in \mathcal{P}_{t}} \nabla_{\Cemb} G_\theta(\Qemb, \Cemb)[i] \right),
\end{split}
\end{equation}
where the summation aggregates the gradient values over all occurrences of the target token in $w$.

Finally, let $V$ be the set of identifier-compatible tokens within the model's vocabulary. We identify the optimal replacement $t^*$ by searching for the candidate that maximizes the influence:
\begin{equation*}
    t^* =\operatorname*{argmax}_{t'\in V}\ \ \textit{influence}(t, t', \mathcal{P}_{t}).
\end{equation*}

Crucially, under the first-order approximation, the perturbations operate in orthogonal subspaces because the position sets $\mathcal{P}_{t}$ for distinct tokens are disjoint. The independence allows us to optimize each token replacement via a greedy approach without requiring combinatorial optimization.

% To ensure the attack preserves the code's execution logic, we enforce two renaming constraints:
% \begin{itemize}[nosep,noitemsep, leftmargin=*]
%     \item \textbf{Consistency}: Every occurrence of a specific identifier $w$ is replaced by the same new identifier $w'$.
%     \item \textbf{Uniqueness}: Distinct identifiers are replaced by distinct new identifiers.
% \end{itemize}
% Implementation details for these constraints are provided in Appendix~\ref{appendix: implementation_details}.

To ensure the attack preserves the code's execution logic, we also enforce two renaming constraints: 1) \textbf{Consistency}: Every occurrence of a specific identifier $w$ is replaced by the same new identifier $w'$; and
2) \textbf{Uniqueness}: Distinct identifiers are replaced by distinct new identifiers.
Implementation details for these constraints are provided in Appendix~\ref{appendix: implementation_details}.

\subsection{Identifier Style Constraints}
\label{ssec: preserving_identifier_format}

As illustrated in Figure~\ref{fig:method_illustration}, naive application of gradient-based optimization often yields long, unintelligible identifiers. Such anomalies make the adversarial code easily detectable by both human programmers and coding agents. To mitigate this, we enforce style-consistency constraints designed to preserve common naming conventions like \texttt{camelCase} and \texttt{snake\_case}. We therefore filter the search space to ensure that valid replacements preserve the original style by including at least the same number of underscores or uppercase letters. Additionally, tokens serving solely as structural separators (e.g., standalone underscores) are irreplaceable. Empirically, we find that enforcing these constraints improves the visual naturalness and plausibility of the code with a marginal reduction in attack efficacy.

\subsection{Query Token Similarity}
\label{ssec: incorporating_query_similarity}

We observe that code search models often exhibit a bias towards snippets that share explicit textual overlap with the target query. To exploit this property, we augment the gradient-based selection objective with a term that rewards proximity to query tokens. Specifically, when considering a candidate replacement token $t'$, we also calculate its similarity to the most aligned token within the query $Q$. With the refined objective, we find the optimal replacement token $t^*$ as
\begin{equation}
\label{eq: token_replacement_final}
    t^* = \operatorname*{argmax}_{t'\in V} \Big[ \textit{influence}(t, t', \mathcal{P}_{t}) + \alpha \max_{q \in Q} (\mathbf{e}_q^\top \mathbf{e}_{t'}) \Big],
\end{equation}
where $\mathbf{e}_{q}$ and $\mathbf{e}_{t'}$ represent the embeddings of a query token $q$ and the candidate replacement $t'$, respectively. Here, the search space V is also restricted to candidates that satisfy the style constraints detailed in Section~\ref{ssec: preserving_identifier_format}. The term $\max_{q \in Q} (\mathbf{e}_q^\top \mathbf{e}_{t'})$ acts as a guidance signal, encouraging the selection of tokens that are similar in embedding space to tokens in the query. We introduce a hyperparameter $\alpha$ to balance the gradient-based influence and this query token similarity regularizer. Empirically, we set $\alpha = 0.1$; detailed ablation studies justifying this choice are provided in Appendix~\ref{appendix: ablation_studies}.

\subsection{Attack Transfer}
\label{ssec: attack_transfer_method}
The proposed attack framework requires white-box access to model parameters and sufficient computational resources to back-propagate gradients and compute token similarities. These prerequisites may not be met in real-world scenarios, particularly when targeting large-scale or proprietary black-box code search systems. 

To address this, we introduce Attack Transfer. Consider a target model with parameters $\theta^*$, which is inaccessible or too computationally expensive to attack directly. We can instead generate the adversarial snippet $C'=\AAttack(Q,\theta)$ using a smaller, accessible surrogate model with parameters $\theta$. Our method relies on the observation that adversarial examples are usually model-agnostic: that is, an optimization that maximizes $\Sim_{\theta}(Q, C')$ is likely to also have high $\Sim_{\theta^*}(Q, C')$. The cross-model transferability allows us to attack complex victim models by optimizing solely on the surrogate.
\section{Experiments and Results}
\label{sec: experiments_and_results}

In this section, we evaluate our adversarial attack across embedding models, code search benchmarks, and LLM-based retrieval. We first quantify the attack's effectiveness and cross-model transferability (Section~\ref{ssec: effectiveness_and_transfer}) and compare it with existing baselines (Section~\ref{ssec: comparison_with_baselines}). We then examine the attack's impact on code search benchmarks in both per-query and shared-corpus settings (Sections~\ref{ssec: application_on_code_search_benchmarks} and~\ref{appendix: shared_corpus_benchmark_attack}) and evaluate the attack's transferability to LLM-based retrieval (Section~\ref{ssec: effectiveness_on_llms}). Finally, we investigate finetuning-based defenses (Section~\ref{ssec: robustness_ft}) and include an ablation study about each individual attack component (Section~\ref{ssec: ablation_studies}).

Unless otherwise noted, adversarial attacks are applied to \textit{(query, code)} pairs. For each pair, we execute our adversarial attack for 5 iterations, creating a candidate set that includes each iteration's output and the original code. From the set, we select the snippet with the highest similarity score to the target query as the final adversarial example. In other words, when no perturbation yields a positive shift in similarity, the original code snippet is preserved, resulting in zero change. The shared-corpus experiment additionally considers adversarial snippets optimized for multiple queries at once. The design choices are justified in Appendix~\ref{appendix: ablation_studies}.

\textbf{Models.} We generate attacks using two surrogate embedding models, \texttt{CodeT5+}~\citep{codet5+} and \texttt{OASIS}~\citep{oasis}, and evaluate transfer to \texttt{Nomic-embed-code}~\citep{nomic-embed-code}, \texttt{Voyage-code-3}~\citep{voyage_code_3}, and the generative models \texttt{GPT-5.4-mini} and \texttt{Gemini-3.1-Pro}. Model details are provided in Appendix~\ref{appendix: model_and_dataset_details}.

\textbf{Datasets.} We use CosQA~\citep{cosqa} and CLARC~\citep{clarc} to evaluate attack effectiveness, transferability, and code-search degradation, and RepoQA~\citep{repoqa} to measure the impact on generative-model retrieval. HumanEval-X~\citep{humaneval_x} is used only for the cross-language analysis in the appendix. Dataset statistics are provided in Table~\ref{tab:dataset_overview}.

% --- Table 1: Models ---
% \begin{table}[t]
%     \centering
%     \small
%     \setlength{\tabcolsep}{3pt}
%     \caption{Models Used in the Experiments.}
%     \label{tab:model_comparison}
%     \begin{tabular}{@{}lccc@{}}
%         \toprule
%         Model & \# Parameters & \makecell{Vocabulary \\ Size} & \makecell{\# Tokens valid \\ for identifiers} \\
%         \midrule
%         CodeT5+ & 110M & 32,103 & 29,881 \\
%         OASIS & 1.54B & 151,665 & 74,194 \\
%         Nomic-embed-code & 7.07B & 151,665 & 74,194 \\
%         Voyage-code-3 & - & 151,665 & 74,194 \\
%         GPT-4o & - & 200,019 & -\\
%         Llama3-70B & 70.6B & 128,000 & -\\
%         \bottomrule
%     \end{tabular}
% \end{table}% % --- Table 2: Datasets ---
% \begin{table}[t]
%     \centering
%     \small
%     \setlength{\tabcolsep}{3.5pt}
%     \caption{Datasets Used in the Experiments.}
%     \label{tab:dataset_overview}
%     \begin{tabular}{@{}lccl@{}}
%         \toprule
%         Dataset & \# Queries & \makecell{\# Code Snippets} & \makecell{Programming \\ Languages} \\
%         \midrule
%         CosQA & 500 & 500 & Python \\
%         CLARC & 526 & 526 & C++ \\
%         RepoQA & 200 & 2,232 & \makecell[l]{Python, C++} \\
%         HumanEval-X & 820 & 820 & \makecell[l]{Python, C++, Java, \\ Javascript, Go} \\

%         \bottomrule
%     \end{tabular}
% \end{table}

\subsection{Effectiveness and Transferability} 
\label{ssec: effectiveness_and_transfer}

\begin{table*}[t]
\centering
\caption{Similarity Changes and Transferability. The \hl{shaded rows} indicate identical Surrogate and Eval models. The substantial similarity increases ($\Delta$ Sim) observed across all rows demonstrate both high attack effectiveness and robust transferability to diverse Eval models. High precision and strong Pearson ($r$) and Spearman ($\rho$) correlation coefficients further confirm that similarity improvements reliably transfer. $\Delta$ Sim, $r$, and $\rho$ are scaled by 100.}
\label{tab:effectiveness_and_transfer}
\scriptsize
% Setup siunitx for this table
\sisetup{
  separate-uncertainty = true,   % v2 equivalent of uncertainty-mode=separate
  table-align-uncertainty = false
}
\setlength{\tabcolsep}{8pt} % Adjust column spacing slightly for fit
\begin{tabular}{lll
                    S[table-format=2.2(2)] % Sim Change: XX.XX +- Y.YY
                    c                     % Pos Count
                    c                     % Neg Count
                    c                     % Precision
                    c                     % Pearson
                    c                     % Spearman
                    }
\toprule
Dataset & Surrogate Model & Eval Model & {$\Delta$ Sim.} & {\makecell{Improved \\ Counts}} & {\makecell{Unimproved \\ Counts}} & Precision & $r$ & $\rho$ \\
\midrule

\multirow{8}{*}{\makecell{CosQA\\(Python)}} & \multirow{4}{*}{CodeT5+}
 & \cellcolor{tablerowhighlight}CodeT5+ & \cellcolor{tablerowhighlight}38.86 \pm 10.44 & \cellcolor{tablerowhighlight}9899 & \cellcolor{tablerowhighlight}101 & \cellcolor{tablerowhighlight}- & \cellcolor{tablerowhighlight}- & \cellcolor{tablerowhighlight}- \\
 & & OASIS & 18.15 \pm 6.03 & 9898 & 102 & 99.99 & 63.37 & 58.69 \\
 & & Nomic-embed-code & 39.25 \pm 10.89 & 9895 & 105 & 99.96 & 62.34 & 55.89 \\
 & & Voyage-code-3 & 27.13 \pm 8.96 & 9896 & 104 & 99.97 & 64.24 & 59.93 \\
 \cmidrule{2-9}
 & \multirow{4}{*}{OASIS}
 & CodeT5+ & 23.23 \pm 12.57 & 9670 & 330 & 98.01 & 75.56 & 75.49 \\
 & & \cellcolor{tablerowhighlight}OASIS & \cellcolor{tablerowhighlight}13.68 \pm 6.89 & \cellcolor{tablerowhighlight}9866 & \cellcolor{tablerowhighlight}134 & \cellcolor{tablerowhighlight}- & \cellcolor{tablerowhighlight}- & \cellcolor{tablerowhighlight}- \\
 & & Nomic-embed-code & 28.37 \pm 12.91 & 9803 & 197 & 99.36 & 84.57 & 83.87 \\
 & & Voyage-code-3 & 19.04 \pm 10.08 & 9810 & 190 & 99.37 & 88.40 & 88.09 \\
\midrule

\multirow{8}{*}{\makecell{CLARC\\(C++)}}& \multirow{4}{*}{CodeT5+}
 & \cellcolor{tablerowhighlight}CodeT5+ & \cellcolor{tablerowhighlight}26.60 \pm 9.13 & \cellcolor{tablerowhighlight}9866 & \cellcolor{tablerowhighlight}134 & \cellcolor{tablerowhighlight}- & \cellcolor{tablerowhighlight}- & \cellcolor{tablerowhighlight}- \\
 & & OASIS & 13.10 \pm 5.75 & 9825 & 175 & 99.64 & 60.70 & 55.61 \\
 & & Nomic-embed-code & 24.66 \pm 9.94 & 9797 & 203 & 99.64 & 62.53 & 56.34 \\
 & & Voyage-code-3 & 13.96 \pm 6.99 & 9854 & 146 & 99.67 & 58.49 & 54.53 \\
\cmidrule{2-9}
 & \multirow{4}{*}{OASIS}
 & CodeT5+        & 12.78 \pm 8.79 & 9360 & 640 & 95.72 & 69.21 & 70.79 \\
 & & \cellcolor{tablerowhighlight}OASIS & \cellcolor{tablerowhighlight}9.21 \pm 5.93 & \cellcolor{tablerowhighlight}9741 & \cellcolor{tablerowhighlight}259 & \cellcolor{tablerowhighlight}- & \cellcolor{tablerowhighlight}- & \cellcolor{tablerowhighlight}- \\
 & & Nomic-embed-code & 15.79 \pm 9.89 & 9607 & 393 & 98.64 & 87.22 & 88.38 \\
 & & Voyage-code-3 & 8.80 \pm 6.55 & 9510 & 490 & 96.57 & 85.85 & 86.74 \\
\bottomrule
\end{tabular}%

\end{table*}

\begin{figure}[t]
    \centering
    % \Description{Histograms showing the distribution of cosine similarity changes after the attack on CosQA, with separate subplots for different evaluation models; most changes are positive and negative shifts are comparatively small.}
    \includegraphics[width=0.45\textwidth]{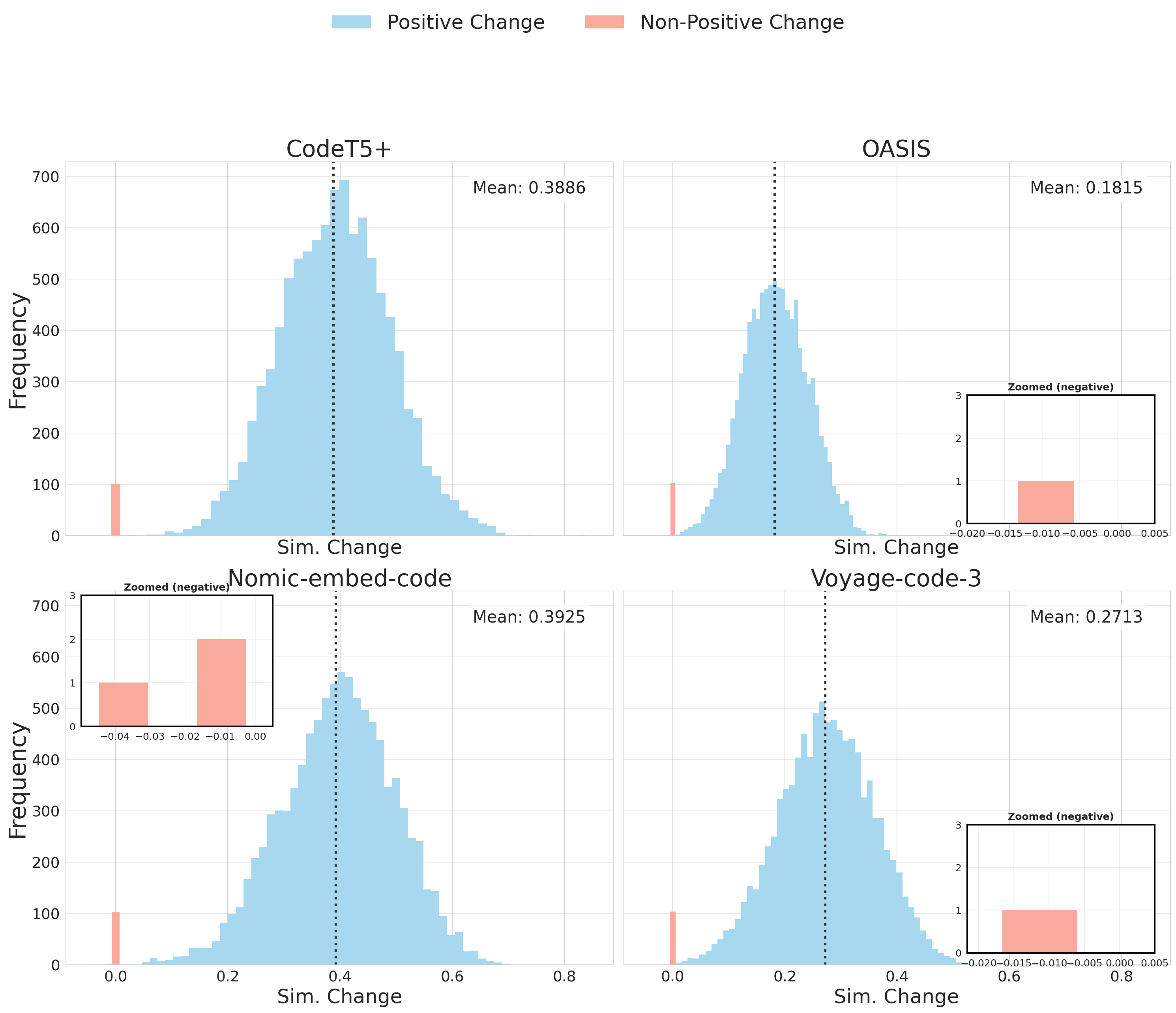}
    \caption{Distribution of Code Similarity Changes on CosQA. The Surrogate Model is \texttt{CodeT5+}, and the Eval Models are labeled at the top of each subplot. Negative similarity changes are smaller in magnitude than positive ones.}
    \label{fig:code_sim_change}
\end{figure}

We evaluate attack effectiveness on 20,000 \textit{(query, code)} pairs sampled from CosQA~\citep{cosqa} and CLARC~\citep{clarc}. Attacks are generated on \texttt{CodeT5+} or \texttt{OASIS} as Surrogate Models and then evaluated on multiple Eval Models in Table~\ref{tab:effectiveness_and_transfer}. Metric details are provided in Appendix~\ref{appendix: details_about_metrics}.

\textbf{Effectiveness on Surrogate Models.}
The shaded rows show that the attack increases query-code similarity for over 97\% of examples on the Surrogate Model. The gains are generally larger for \texttt{CodeT5+} than for \texttt{OASIS}, which may reflect the denser embedding space of \texttt{OASIS} (Appendix~\ref{appendix:density_of_embedding}) and its robustness-oriented training~\citep{oasis}. The attack is also stronger on CosQA than on CLARC, suggesting that dataset or language characteristics affect attack magnitude.

\textbf{Attack Transferability.}
The unshaded rows show strong transfer to diverse Eval Models, including robustness-enhanced, finetuned, and closed-source systems. Across Surrogate/Eval pairs, the attack increases similarity in over 95\% of cases, with high precision and positive Pearson/Spearman correlations indicating that surrogate gains generally predict transfer gains. When transfer fails, the negative shifts are small compared to successful gains (Figure~\ref{fig:code_sim_change} and Appendix~\ref{appendix: distribution_of_code_similarity_changes}).

Tokenizer differences do not block transferability. Although \texttt{CodeT5+} has a much smaller vocabulary than \texttt{Nomic-embed-code} or \texttt{Voyage-code-3}, it still transfers effectively to these larger models. Overall, \texttt{CodeT5+}-based attacks tend to produce larger $\Delta$ Sim, while \texttt{OASIS}-based attacks yield more consistent correlations.

Beyond cross-model transfer, the attack also transfers across queries sharing the same intents. As detailed in Appendix~\ref{ssec: query_transferability}, adversarial snippets optimized for one query remain effective for semantically equivalent paraphrases, suggesting that the attack targets the underlying intent rather than the specific target query string.

% \st{To further verify the language-agnostic nature of our method, we design an experiment to validate the effectiveness of our adversarial attack between queries and functionally equivalent implementations in Python, C++, Java, JavaScript, and Go. Detailed results confirming consistent performance are provided in Appendix.}
% \kw{Already moved to Section 4.4}

\subsection{Comparison with Baselines} 
\label{ssec: comparison_with_baselines}

We compare Attack Transfer with two baseline attacks on 1,000 \textit{(query, code)} pairs from CosQA. In the white-box setting, we transfer attacks generated on \texttt{CodeT5+} to \texttt{OASIS} and compare against DAMP~\cite{DAMP} applied directly to \texttt{OASIS}. In the black-box setting, we transfer attacks from \texttt{CodeT5+} to \texttt{Voyage-code-3} and compare against CodeAttack~\citep{codeattack}. Since DAMP and CodeAttack were originally designed for code classification, we adapt them to the code search setting.

\begin{table}[t]
    \centering
    \caption{Comparison of Attack Transfer against baselines. Our method achieves stronger attack effectiveness with substantially lower computational or query cost. GPU hours are calculated on a NVIDIA L40 ADA 48GB GPU.}
    % \vspace{-0.2em}
    \label{tab:baseline_comparison}
    \scriptsize
    \setlength{\tabcolsep}{10pt}
\begin{tabular}{@{}l r@{ $\pm$ }l cr@{}} 
    \toprule
    Method & \multicolumn{2}{c}{$\Delta$ Sim.} & {\makecell{Improved Counts}} & Efficiency \\
    \midrule
    \multicolumn{5}{@{}l}{\textit{\textbf{White Box}}} \\ 
    DAMP & 10.45 & 5.48 & 978 & 81 mins \\
    Ours & 18.66 & 5.52 & 999 & 47 mins \\
    \midrule
    \multicolumn{5}{@{}l}{\textit{\textbf{Black Box}}} \\ 
    CodeAttack & 5.18 & 4.18 & 907 & ~10.8m Calls \\
    Ours & 28.03 & 8.29 & 1,000 & 1k Calls \\
    \bottomrule
    \end{tabular}%

\end{table}

Table~\ref{tab:baseline_comparison} shows that our transferred attack yields stronger effectiveness and better efficiency. In the white-box setting, Attack Transfer produces a larger $\Delta$ Sim. than DAMP while using less GPU time. In the black-box setting, the gap is larger: our transferred attack achieves about a 5$\times$ higher $\Delta$ Sim. than CodeAttack while reducing the required queries from $\sim$10.8M API calls to 1k calls. The baseline attacks also transfer less effectively across models, as detailed in Appendix~\ref{appendix: baseline_transferability}.

% \kw{Add a section of the transferability baseline.}

\subsection{Application on Code Search Benchmarks}
\label{ssec: application_on_code_search_benchmarks}

In this experiment, \texttt{CodeT5+} is used as the Surrogate Model. For each query in the CosQA and CLARC datasets, we select 10\%\footnote{The results when other ratios of the corpus are attacked are available in Appendix~\ref{appendix: application_on_benchmark_appendix}.} of irrelevant code snippets and apply our attack to maximize their similarity to the query. These adversarially modified snippets replace the original ones in the candidate pools. Eval Models (\texttt{CodeT5+}, \texttt{OASIS}, \texttt{Nomic-embed-code}, and \texttt{Voyage-code-3}) then embed the queries and the modified pools, and rerank the code snippets by similarity. The retrieval metrics are measured before and after the adversarial attack.

As shown in Table~\ref{tab:application_on_cosqa_summary}, replacing only 10\% of irrelevant candidates with adversarial versions results in sharp drops across all models and retrieval metrics. Before the attack, models like \texttt{OASIS}, \texttt{Nomic-embed-code}, and \texttt{Voyage-code-3} achieve strong R@5 ($>$90\%) and MRR (74-87\%) scores, suggesting the task is nearly ``solved.'' However, after the attack, MRR drops to single digits, and R@5 drops to below 15\% for all Eval Models, revealing that high benchmark scores do not guarantee robustness against adversarial manipulation.

As expected, \texttt{CodeT5+}, the Surrogate Model, suffers severe degradation. Yet the attack also transfers effectively to other models. \texttt{OASIS} remains vulnerable, despite being trained against hard negatives with similar keywords on an augmented dataset. \texttt{Nomic-embed-code}'s substantial performance decrease indicates even large CLMs can overly depend on lexical features like identifiers, suggesting that scale-up does not naturally yield greater robustness. While \texttt{Voyage-code-3} is the most resilient when allowed to retrieve more candidates (e.g., Recall@20), it still suffers from a drastic drop in MRR and Recall@1, confirming that the attack effectively buries the ground truth beneath adversarial noise.

\begin{table}[t]
\centering
\footnotesize
% 1. Setup rounding rules locally for this table
\sisetup{
    round-mode          = places,
    round-precision     = 2,
    table-format        = -3.2,  % <--- The minus sign here is the key
}

\caption{Code search degradation under a 10\% adversarial attack on CosQA using \texttt{CodeT5+} as the surrogate model. The attack causes large drops across retrieval metrics and Eval Models. Full metrics are reported in Appendix~\ref{appendix: application_on_benchmark_appendix}.}
\label{tab:application_on_cosqa_summary}
\scriptsize
\setlength{\tabcolsep}{1pt}
\begin{tabular}{@{} l *{6}{S} @{}}
\toprule
& \multicolumn{3}{c}{CodeT5+} & \multicolumn{3}{c}{Nomic-embed-code} \\
\cmidrule(lr){2-4} \cmidrule(l){5-7}
Setting & \multicolumn{1}{r}{MRR} & \multicolumn{1}{r}{NDCG} & \multicolumn{1}{r}{R@1} & \multicolumn{1}{r}{MRR} & \multicolumn{1}{r}{NDCG} & \multicolumn{1}{r}{R@1} \\
\midrule
Original     & 74.08 & 78.52 & 64 & 83.03 & 86.55 & 73.8 \\
Adversarial  & 1.79 & 2.46 & 1.2 & 5.95 & 8.13 & 3 \\
\cellcolor{tablerowhighlight}$\Delta$ & \cellcolor{tablerowhighlight}\color{red}-72.29 & \cellcolor{tablerowhighlight}\color{red}-76.06 & \cellcolor{tablerowhighlight}\color{red}-62.8 & \cellcolor{tablerowhighlight}\color{red}-77.07 & \cellcolor{tablerowhighlight}\color{red}-78.42 & \cellcolor{tablerowhighlight}\color{red}-70.8 \\
\midrule
& \multicolumn{3}{c}{OASIS} & \multicolumn{3}{c}{Voyage-code-3} \\
\cmidrule(lr){2-4} \cmidrule(l){5-7}
Setting & \multicolumn{1}{r}{MRR} & \multicolumn{1}{r}{NDCG} & \multicolumn{1}{r}{R@1} & \multicolumn{1}{r}{MRR} & \multicolumn{1}{r}{NDCG} & \multicolumn{1}{r}{R@1} \\
\midrule
Original     & 80.57 & 84.73 & 71 & 87.06 & 89.9 & 79.4 \\
Adversarial  & 5.99 & 8.86 & 2.6 & 9.52 & 13.27 & 4.4 \\
\cellcolor{tablerowhighlight}$\Delta$ & \cellcolor{tablerowhighlight}\color{red}-74.58 & \cellcolor{tablerowhighlight}\color{red}-75.87 & \cellcolor{tablerowhighlight}\color{red}-68.4 & \cellcolor{tablerowhighlight}\color{red}-77.54 & \cellcolor{tablerowhighlight}\color{red}-76.63 & \cellcolor{tablerowhighlight}\color{red}-75 \\
\bottomrule
\end{tabular}%
\vspace{-1em}
\end{table}

\subsection{Shared Corpus Attack}
\label{appendix: shared_corpus_benchmark_attack}

% \kw{Moved from Appendix}

The experiments in Section~\ref{ssec: application_on_code_search_benchmarks} attack a separate candidate pool for each query. We now consider the more constrained setting in which all queries are retrieved against a single shared corpus, so that every adversarial snippet is simultaneously visible to all queries. We sample 20 queries and construct a corpus of 500 code snippets, consisting of the 20 ground-truth snippets and 480 irrelevant ones. All the queries and the code snippets are from CosQA.

We first randomly select 50 of the 480 irrelevant snippets and attack them with \texttt{CodeT5+} as the Surrogate Model, targeting each of the 20 queries. For every query, we retain only the single most effective adversarial snippet, i.e., the one with the highest \texttt{CodeT5+} similarity to that query. Inserting these snippets into the corpus yields the ``20 attacked snippets'' setting, in which each query is targeted by exactly one adversarial snippet.

To further reduce the number of attacked snippets, we extend the optimization objective in Equation~\ref{eq: token_replacement_final}, which targets a single query, by aggregating the influence and query-similarity terms over a set of queries. A single adversarial snippet can therefore be optimized against several queries at once. Using this multi-query objective, we generate snippets that each target 2 or 4 queries, producing the ``10 attacked snippets'' and ``5 attacked snippets'' settings, respectively.

Table~\ref{tab:attack_results} reports retrieval performance under the shared-corpus attack. Injecting 20 adversarial snippets, i.e., 4\% of the corpus, sharply reduces MRR and NDCG for all Eval Models. Recall@5 is largely unaffected: with one adversarial snippet per query, the ground truth is displaced from the top rank but rarely beyond the top five. The 10- and 5-snippet settings confirm that the multi-query objective remains effective, degrading MRR and NDCG with only 1-2\% of the corpus attacked, though the effect weakens as each snippet is shared among more queries, revealing a trade-off between the number of attacked snippets and per-query attack strength.

\begin{table}[t]
\centering
\caption{Retrieval performance under the shared-corpus attack across different numbers of injected snippets. Values in parentheses denote the absolute drop ($\textcolor{red}{\downarrow}$) relative to the unattacked baseline; ($\rightarrow$) indicates no change.}
\label{tab:attack_results}
\scriptsize
% requires \usepackage{xcolor} and \usepackage{etoolbox} in the preamble
\newcommand{\cell}[2]{\makebox[2.4em][r]{#1}\makebox[3.4em][l]{\ifstrempty{#2}{\ \textcolor{black}{($\rightarrow$)}}{\ \textcolor{red}{($\downarrow$\ #2)}}}}
\setlength{\tabcolsep}{12pt}
\begin{tabular}{@{}lccc@{}}
\toprule
Model & MRR & NDCG & Recall@5 \\
\midrule
% \multicolumn{4}{l}{\textbf{No attack}} \\
% CodeT5+ & \cell{79.1}{} & \cell{84.2}{} & \cell{95.0}{} \\
% OASIS   & \cell{78.9}{} & \cell{83.9}{} & \cell{90.0}{} \\
% Nomic   & \cell{81.2}{} & \cell{84.8}{} & \cell{95.0}{} \\
% \midrule
\multicolumn{4}{@{}l}{\textbf{20 attacked snippets}} \\
CodeT5+ & \cell{40.8}{38.2} & \cell{54.6}{29.6} & \cell{90.0}{5.0} \\
OASIS   & \cell{67.0}{11.9} & \cell{74.9}{9.0}  & \cell{90.0}{} \\
Nomic   & \cell{66.8}{14.4} & \cell{74.0}{10.8} & \cell{95.0}{} \\
\midrule
\multicolumn{4}{@{}l}{\textbf{10 attacked snippets}} \\
CodeT5+ & \cell{69.8}{9.3} & \cell{77.2}{7.0} & \cell{90.0}{5.0} \\
OASIS   & \cell{73.9}{5.0} & \cell{80.1}{3.8} & \cell{90.0}{} \\
Nomic   & \cell{76.2}{5.1} & \cell{81.0}{3.8} & \cell{95.0}{} \\
\midrule
\multicolumn{4}{@{}l}{\textbf{5 attacked snippets}} \\
CodeT5+ & \cell{74.7}{4.4} & \cell{80.8}{3.4} & \cell{90.0}{5.0} \\
OASIS   & \cell{74.5}{4.3} & \cell{80.5}{3.4} & \cell{90.0}{} \\
Nomic   & \cell{75.4}{5.8} & \cell{80.4}{4.4} & \cell{95.0}{} \\
\bottomrule
\end{tabular}
\end{table}

\subsection{Effectiveness on LLMs}
\label{ssec: effectiveness_on_llms}

% \begin{table}[t]
% \centering
% \caption{Impact of adversarial attacks on LLMs evaluated on RepoQA. We compare accuracy in the Original setting against the Full Attack setting and its ablations: w/o Grad. removes gradient information, w/o Style removes style preservation, and w/o QTS removes query-token similarity. The results reveal a severe reduction in accuracy for both \texttt{GPT-4o} and \texttt{Llama3-70B}, confirming their susceptibility to our attack.}
% \label{tab:application_on_LLM}
% \small
% \setlength{\tabcolsep}{3pt}
% \begin{tabular}{llccc}
% \toprule
% Model & Setting & Python & C++ & Java \\
% \midrule
% \multirow{5}{*}{GPT-4o} 
%     & Original               & 95\% & 80\% & 96\% \\
%     \cmidrule{2-5}
%     & Full Attack            & 72\% & 51\% & 67\% \\
%     & \ \ \ \  w/o Grad.    & 81\% & 66\% & 79\% \\
%     & \ \ \ \  w/o Style    & 89\% & 72\% & 90\% \\
%     & \ \ \ \  w/o QTS      & 91\% & 74\% & 93\% \\
% \midrule
% \multirow{5}{*}{Llama3-70B} 
%     & Original               & 83\% & 70\% & 86\% \\
%     \cmidrule{2-5}

%     & Full Attack            & 51\% & 39\% & 56\% \\
%     & \ \ \ \  w/o Grad.    & 62\% & 57\% & 61\% \\
%     & \ \ \ \  w/o Style    & 79\% & 64\% & 77\% \\
%     & \ \ \ \  w/o QTS      & 81\% & 63\% & 82\% \\
% \bottomrule
% \end{tabular}
% \end{table}

\begin{table}[t]
\centering
\caption{Impact of adversarial attacks on LLMs evaluated on RepoQA. All reported statistics are accuracy percentages. We compare accuracy in the Original setting against the Full Attack setting and its ablations: w/o Grad. removes gradient information, w/o Style removes style preservation, and w/o QTS removes query-token similarity. The results reveal a severe reduction in accuracy for both \texttt{GPT-5.4-mini} and \texttt{Gemini-3.1-Pro}, confirming their susceptibility to our attack.}
\label{tab:application_on_LLM}
\scriptsize
\setlength{\tabcolsep}{3pt}
\begin{tabular}{llccccc}
\toprule
Model & Setting & Python & C++ & Java & Rust & TypeScript \\
\midrule
\multirow{5}{*}{GPT-5.4-mini} 
    & Original               & 96 & 92 & 94 & 92 & 99 \\
    \cmidrule{2-7}
    & Full Attack            & 81 & 70 & 73 & 78 & 83 \\
    & \ \ \ \ w/o Grad.      & 87 & 79 & 81 & 86 & 86 \\
    & \ \ \ \ w/o Style      & 93 & 87 & 88 & 90 & 94 \\
    & \ \ \ \ w/o QTS        & 89 & 85 & 85 & 90 & 93 \\
\midrule
\multirow{5}{*}{Gemini-3.1-Pro} 
    & Original               & 95 & 92 & 95 & 95 & 98 \\
    \cmidrule{2-7}
    & Full Attack            & 79 & 72 & 77 & 80 & 82 \\
    & \ \ \ \ w/o Grad.      & 88 & 80 & 79 & 88 & 85 \\
    & \ \ \ \ w/o Style      & 91 & 90 & 89 & 92 & 95 \\
    & \ \ \ \ w/o QTS        & 87 & 81 & 82 & 90 & 93 \\
\bottomrule
\end{tabular}
\vspace{-1em}
\end{table}

We further test whether the attack transfers from embedding-based code search to LLM-based repository retrieval. Following RepoQA~\cite{repoqa}, each LLM receives a detailed natural-language query and a candidate set, then selects the matching function; accuracy is the evaluation metric. For each query, we replace only \textbf{two} irrelevant candidates (less than 10\% of the input tokens) with adversarial snippets generated using \texttt{CodeT5+}.

As shown in Table~\ref{tab:application_on_LLM}, this small perturbation can transfer to LLMs and consistently reduce accuracy for both \texttt{GPT-5.4-mini} and \texttt{Gemini-3.1-Pro} across Python, C++, and Java. The ablation rows indicate that each component contributes to the LLM transfer effect; we discuss these components in more detail in Section~\ref{ssec: ablation_studies}.

\subsection{Defense via Robust Finetuning}
\label{ssec: robustness_ft}

\begin{table}[t]
\centering
\scriptsize
\caption{Retrieval and robustness trade-off after finetuning. PT-CodeT5+ is generated on pretrained \texttt{CodeT5+}; Trans-OASIS is transferred from \texttt{OASIS}; FT-CodeT5+ is generated directly on the evaluated finetuned \texttt{CodeT5+} checkpoint.}
%All robustness values are $\Delta$ Sim.}
\label{tab:robust_ft_summary}
\setlength{\tabcolsep}{1.2pt}
\begin{tabular}{@{}lccc@{\hspace{2pt}}ccc@{}}
\toprule
\multirow{2}{*}{Checkpoint} & \multicolumn{3}{c}{Retrieval} & \multicolumn{3}{c}{Robustness ($\Delta$ Sim.)} \\
\cmidrule(lr){2-4} \cmidrule(lr){5-7}
& MRR & NDCG & R@5 & \makecell{PT-CodeT5+\\Attack} & \makecell{OASIS\\Transfer} & \makecell{FT-CodeT5+\\Attack} \\
\midrule
Pretrain Base & 72.4 & 77.3 & 88.0 & 39.90 & 22.25 & -- \\
Robust-Only FT & 34.1 & 37.6 & 42.6 & -22.70 & -15.70 & 7.14 \\
Mixed FT & 72.3 & 77.6 & 89.4 & 4.00   & 4.42   & 34.70 \\
\bottomrule
\end{tabular}
\end{table}

We also examine whether finetuning can defend against our attack.\footnote{Details of Robust-Only FT and Mixed FT are provided in Appendix~\ref{appendix: robustness_ft}.} As summarized in Table~\ref{tab:robust_ft_summary}, the first key finding is that Robust-Only FT can mitigate the attack's influence, driving static attack-induced $\Delta$ Sim. below zero and substantially reducing the white-box attack effect. However, this robustness comes with a prohibitive retrieval cost: the main retrieval metrics fall to roughly half of the pretrained baseline, with R@5 dropping from 88.0 to 42.6. The second key finding is that Mixed FT appears to balance retrieval quality and static robustness better, preserving baseline search performance while reducing the impact of static attacks. Nevertheless, this balance breaks under an adaptive white-box threat model: when the attack is generated directly on the finetuned checkpoint, Mixed FT remains highly vulnerable. Moreover, although not shown in the main table, Appendix~\ref{appendix: robustness_ft} shows that attacks on the Mixed FT checkpoint also transfer to other Eval Models. Overall, standard adversarial finetuning is therefore insufficient as a defense: robust-only training sacrifices retrieval utility, while mixed training preserves utility but leaves an effective adaptive and transferable attack surface.

\subsection{Ablation Studies}
\label{ssec: ablation_studies}

% \todo{Extend more, the current discussion is too concise}

To isolate the contribution of each component described in Sections~\ref{ssec: gradient_based_method}, \ref{ssec: preserving_identifier_format}, and \ref{ssec: incorporating_query_similarity}, we remove each component from the full method and evaluate both embedding-level similarity on 10,000 CosQA pairs (Table~\ref{tab:ablation_study}) and downstream LLM retrieval accuracy on RepoQA (Table~\ref{tab:application_on_LLM}). In the experiment, CodeT5+ is still used as the surrogate model. Other ablation studies on the selection of the adversarial code snippet, the number of iterations, and the choice of the hyperparameter $\alpha$ are provided in Appendix~\ref{appendix: ablation_studies}.

\begin{table}[t]
\centering
\caption{Ablation study quantifying the contribution of specific components. Improved Counts denote the number of pairs with increased similarity after the attack.}
\label{tab:ablation_study}
\scriptsize
\setlength{\tabcolsep}{3pt}
\begin{tabular}{@{} l l c r @{}}
\toprule
Method & Eval Model & $\Delta$ \textbf{Sim.} & Improved Counts \\
\midrule
Full Method            & CodeT5+ & $38.86 \pm 10.44$ & 9,899 \\
                       & Nomic   & $39.25 \pm 10.89$ & 9,895 \\
\midrule
w/o gradient           & CodeT5+ & $30.91 \pm 12.42$ & 9,855 \\
                       & Nomic   & $35.35 \pm 12.66$ & 9,848 \\
\midrule
w/o style preservation & CodeT5+ & $39.48 \pm 10.39$ & 9,899 \\
                       & Nomic   & $39.95 \pm 10.86$ & 9,895 \\
\midrule
w/o query token sim.   & CodeT5+ & $30.93 \pm 10.08$ & 9,893 \\
                       & Nomic   & $22.68 \pm 11.01$ & 9,766 \\
\bottomrule
\end{tabular}
\end{table}

Without gradient guidance, the attack relies solely on query similarity, often replacing identifiers with query tokens by default. When style constraints restrict the search, it selects the most similar style-compliant candidate instead. As shown in Table~\ref{tab:ablation_study}, removing gradient information reduces $\Delta$ Sim. for both \texttt{CodeT5+} and \texttt{Nomic-embed-code}, and the larger standard deviations indicate less stable similarity changes. Since backpropagation consumes most of the computational cost, the gradient-free variant can still serve as a lightweight alternative when resources are limited, but it is consistently weaker than the full method.

Removing style preservation yields negligible, and occasionally positive, changes in raw embedding similarity. This suggests that when an optimal token violates style rules, a style-compliant substitute with comparable embedding efficacy often exists. Further analysis reveals distinct behaviors regarding underscores between evaluation models: in \texttt{CodeT5+}, ``\_'' is typically parsed as an independent token, so replacing it has little influence; in \texttt{OASIS}, underscores are usually treated as token prefixes, and in approximately 90\% of cases, a top-3 candidate token also begins with an underscore. However, Table~\ref{tab:application_on_LLM} shows that style preservation is crucial for generative models. Unconstrained attacks may preserve embedding similarity, but they produce visibly incoherent identifiers that are less convincing to LLMs, causing \texttt{GPT-5.4-mini} and \texttt{Gemini-3.1-Pro} to recover much of their original RepoQA accuracy.

Query token similarity is also essential for transferability. Removing this component causes a significant drop in embedding effectiveness, particularly on \texttt{Nomic-embed-code}. Further analysis indicates that in 83\% of cases, the optimal replacement tokens derived from the full method (Equation~\ref{eq: token_replacement_final}) are among the top-20 candidates with the highest influence. Table~\ref{tab:application_on_LLM} further shows that discarding query token similarity substantially weakens the attack against LLMs. By prioritizing similarity to query tokens, the full method inserts semantically relevant vocabulary rather than only mathematically optimized gradient tokens, improving the naturalness and coherence needed to mislead generative models.

% Table~\ref{tab:ablation_study} shows that gradient information and query-token similarity both improve embedding-level attack strength. Removing gradient guidance lowers $\Delta$ Sim. for both the surrogate and transfer model, while removing query-token similarity causes the largest transfer drop on \texttt{Nomic-embed-code}. In contrast, removing style preservation has little effect on raw similarity, suggesting that style-compliant substitutes often achieve similar embedding gains.

% However, Table~\ref{tab:application_on_LLM} shows that embedding similarity alone does not explain transfer to generative models. Style preservation and query-token similarity are especially important for LLMs: without style preservation, adversarial identifiers become visibly unnatural, and without query-token similarity, replacements lose semantic alignment with the user query. In both cases, \texttt{GPT-5.4-mini} and \texttt{Gemini-3.1-Pro} recover much of their original accuracy across Python, C++, and Java. These results indicate that successful attacks on generative models require not only high embedding influence but also natural, query-relevant perturbations.

\section{Conclusion \& Future Work}

We present a transferable adversarial attack that modifies code identifiers to mislead code search models. The attack is effective across embedding models, retrieval benchmarks, and LLM-based repository retrieval, revealing that current CLMs still rely heavily on lexical cues rather than robust semantic understanding.

Future work should investigate why attacks transfer across models and develop defenses that preserve retrieval utility. Promising directions include functionality-aware contrastive training and incorporating programming-language structure, such as ASTs, to build more semantics-grounded code representations.

% Original version:
% We propose a transferable adversarial attack that modifies code identifiers to mislead code search models, and the attack transfers effectively across multiple models and application scenarios, demonstrating that current CLMs rely heavily on lexical features. The revealed vulnerability highlights the urgent need for more robust and semantics-aware code embedding techniques.
%
% Future research could explore several directions. First, the strong correlation in attack transferability across models warrants further investigation into shared pre-training data or common architectural biases. Second, future work could focus on developing more robust code embedding models through new defense mechanisms or training strategies, such as contrastive learning tailored to functionality-preserving perturbations. Lastly, because the attack exploits the difference between natural and programming languages, simply adapting NLP procedures for CLMs may be inadequate, especially when fine-tuning data might be insufficient to address the unique scenario. Instead, integrating programming language-specific structures, such as Abstract Syntax Trees, could be a more efficient approach toward building more robust and semantically grounded CLMs.

\section*{Acknowledgements}
We thank the anonymous reviewers, ACs, and PCs for their constructive feedback. We also thank Jiaqi Lu for insightful discussions during the early stages of this work. This research was supported in part by the U.S. National Science Foundation (NSF) under grants 2409005 and 2321444. Any opinions, findings, and conclusions in this paper are those of the authors only and do not necessarily reflect the views of our sponsors.

\clearpage
% \section*{Impact Statement}
% This work exposes critical vulnerabilities in code language to motivate improved model robustness. While the proposed adversarial methods possess a dual-use risk, potentially allowing malicious actors to manipulate search rankings, demonstrating these flaws is essential for the community to develop secure, semantically-aware code intelligence tools that rely on deep understanding rather than brittle lexical features.

\bibliography{reference}

%%%%%%%%%%%%%%%%%%%%%%%%%%%%%%%%%%%%%%%%%%%%%%%%%%%%%%%%%%%%%%%%%%%%%%%%%%%%%%%
%%%%%%%%%%%%%%%%%%%%%%%%%%%%%%%%%%%%%%%%%%%%%%%%%%%%%%%%%%%%%%%%%%%%%%%%%%%%%%%
% APPENDIX
%%%%%%%%%%%%%%%%%%%%%%%%%%%%%%%%%%%%%%%%%%%%%%%%%%%%%%%%%%%%%%%%%%%%%%%%%%%%%%%
%%%%%%%%%%%%%%%%%%%%%%%%%%%%%%%%%%%%%%%%%%%%%%%%%%%%%%%%%%%%%%%%%%%%%%%%%%%%%%%
\newpage
\appendix
% \section{Use of LLMs}

% We detail our use of Large Language Models (LLMs) below:

% \begin{itemize}
%     \item \textbf{Experimental Application:} The \texttt{gpt-4o} model was utilized as a component of the Retrieval-Augmented Generation (RAG) pipeline, as presented in Section~\ref{ssec: application_on_rag_systems}. This was the only application of LLMs in this paper.
    
%     \item \textbf{Writing Assistance:} We also employed LLMs to aid in improving the grammar, clarity, and phrasing of the draft during the writing process.
% \end{itemize}

% \section{Limitation}
% \label{appendix: limitations}

% This study has several limitations. Firstly, our presented attacks exclusively target \textit{(query, code)} pairs. While our gradient-based methodology could potentially modify a code snippet to increase its similarity with multiple queries concurrently (i.e., a \textit{(query\_list, code)} input format), such experiments were not conducted due to time constraints.

% Also, although we demonstrated the transferability of adversarial attacks across various code embedding models, we have not identified the underlying reasons for this phenomenon. We hypothesize that shared pretraining data among these models contributes to transferability; however, the number of models tested in this work with publicly available pretraining data was insufficient to draw definitive conclusions.

\section{Limitations}
\label{appendix: limitations}

Although we demonstrate that adversarial attacks can transfer across various code embedding models, the underlying causes of this transferability remain unclear. One possible explanation is that these models share overlapping pretraining data. However, because only a limited number of evaluated models have publicly available pretraining details, we cannot draw a definitive conclusion.

\section{Reproducibility}
\label{appendix:compute_resource}

\begin{table*}[t]
\centering
\caption{Compute resources used to generate 10,000 adversarial examples.}
\label{tab:compute_resource}
\small
\begin{tabular}{lccccc}
\toprule
{Surrogate Model} & \makecell{Total Time} & \makecell{GPU Time} & \makecell{CPU Time} & \makecell{GPU Memory} & \makecell{Token Search Space} \\
\midrule
\textbf{CodeT5+ (110M)} & 8 hours  & $\sim$6.8 hours (85\%) & $\sim$1.2 hours (15\%) & 7GB & 15.1k \\
\textbf{OASIS (1.5B)}   & 12 hours & $\sim$10.6 hours (88\%)& $\sim$1.4 hours (12\%)& 26GB & 36.7k \\
\bottomrule
\end{tabular}
\end{table*}

The experiments described in this paper were conducted on a server equipped with an AMD EPYC Milan 7643 48-core CPU (@2.30GHz), 1TB of RAM, and an NVIDIA L40 Ada 48GB GPU. We used a batch size of 10 \textit{(query, code)} pairs for attacks on \texttt{CodeT5+} and 4 pairs for attacks on \texttt{OASIS}. We observed that larger batch sizes did not substantially change the runtime. The time and GPU memory required for the experiments are reported in Table~\ref{tab:compute_resource}.

To mitigate the risk of potential misuse, we do not publicly release the attack codebase. However, researchers interested in reproducing or extending these results are encouraged to contact the first author to discuss our implementation details.

% For attacks targeting CodeT5+, our experiments consumed approximately 7GB of GPU memory. Processing 100$\times$100 \textit{(query, code)} pairs took about 8 hours for Python, Java, JavaScript, and Go. For C++, the same task required approximately 18 hours, due to its significantly slower code parser.

% For attacks targeting OASIS, experiments utilized around 26GB of GPU memory. The attack duration for 100$\times$100 pairs was about 12 hours for Python, Java, JavaScript, and Go, extending to approximately 21 hours for C++ because of the slower code parser.

\section{Model \& Dataset Details}
\label{appendix: model_and_dataset_details}

\begin{table*}[ht]
    \centering
    \small
    \caption{Models used in the experiments.}
    \label{tab:model_comparison}
    \begin{tabular}{lccc}
        \toprule
        Model & \# Parameters & \makecell{Vocabulary \\ Size} & \makecell{\# Tokens valid \\ for identifiers} \\
        \midrule
        CodeT5+ & 110M & 32,103 & 29,881 \\
        OASIS & 1.54B & 151,665 & 74,194 \\
        Nomic-embed-code & 7.07B & 151,665 & 74,194 \\
        Voyage-code-3 & - & 151,665 & 74,194 \\
        GPT-5.4-mini & - & - & -\\
        Gemini-3.1-Pro & - & - & -\\
        \bottomrule
    \end{tabular}
\end{table*}

% --- Table 2: Datasets ---
\begin{table*}[ht]
    \centering
    \small
    \caption{Datasets used in the experiments.}
    \label{tab:dataset_overview}
    \begin{tabular}{lccl}
        \toprule
        Dataset & \# Queries & \makecell{\# Code Snippets} & Programming Languages \\
        \midrule
        CosQA & 500 & 500 & Python \\
        CLARC & 526 & 526 & C++ \\
        RepoQA & 200 & 2,232 & {Python, C++, Java} \\
        HumanEval-X & 820 & 820 & {Python, C++, Java, JavaScript, Go} \\

        \bottomrule
    \end{tabular}
\end{table*}

\textbf{Models.}
Table~\ref{tab:model_comparison} summarizes the models used in our experiments. \texttt{OASIS}, \texttt{Nomic-embed-code}, and \texttt{Voyage-code-3} use highly similar modified Qwen2 tokenizers~\citep{qwen2}: \texttt{OASIS} and \texttt{Voyage-code-3} share the same tokenizer, while \texttt{Nomic-embed-code} differs only in the index assigned to the \texttt{<EOS>} token. For closed-source models, parameter counts and identifier-token counts are unavailable.

\textbf{Datasets.}
Table~\ref{tab:dataset_overview} summarizes the datasets used in the experiments.

\subsection{Model \& Dataset License}
\label{appendix: model_and_dataset_license}

\begin{itemize}[nosep,noitemsep]
    \item \textbf{CodeT5+}: BSD 3-Clause License \footnote{\url{https://github.com/salesforce/CodeT5?tab=BSD-3-Clause-1-ov-file}}
    \item \textbf{OASIS}: MIT License\footnote{ \url{https://huggingface.co/Kwaipilot/OASIS-code-embedding-1.5B}}
    \item \textbf{Nomic-embed-code}: Apache-2.0 \footnote{\url{https://huggingface.co/nomic-ai/nomic-embed-code}}
    \item \textbf{Voyage-code-3}: The license is unclear; however, we do not include embeddings from \texttt{Voyage-code-3} in our codebase.
    \item \textbf{CosQA}: Apache-2.0\footnote{\url{https://github.com/CoIR-team/coir/blob/main/LICENSE}} (we use CosQA from COIR) 
    \item \textbf{CLARC}: CC-BY-SA 4.0\footnote{\url{https://huggingface.co/datasets/ClarcTeam/CLARC}}
    \item \textbf{RepoQA}: Apache-2.0\footnote{\url{https://github.com/evalplus/repoqa/blob/main/LICENSE}}
    \item \textbf{HumanEval-X}: Apache-2.0\footnote{\url{https://huggingface.co/datasets/THUDM/humaneval-x}}
\end{itemize}

\section{Implementation Details}
\label{appendix: implementation_details}

\subsection{Detailed Constraints}

\newcommand{\tokenbox}[1]{\colorbox{tablerowhighlight}{\strut\texttt{#1}}}

In our adversarial attack, we focus on replacing identifier tokens---specifically, tokens that comprise function, variable, macro, and module names in the code text. Formally, let the original code text be tokenized as $\{C_{t_i}\}_{i=1}^{n}$, and the code text after replacement be tokenized as $\{C'_{t_i}\}_{i=1}^{n}$. For any two strings $A$ and $B$, let $\LCS(A, B)$ denote their longest common substring, and let $\Remove(A, B)$ denote the remaining string after removing string $B$ from string $A$. 

The primary challenge in replacing these identifiers is accounting for tokenization corner cases, where varying punctuation or whitespace is merged with the identifier itself. To address this, we introduce two additional constraints during replacement. We illustrate the intuition behind these constraints using the following original code snippet (top) and its adversarially modified version (bottom):

\begin{lstlisting}[style=pythonpretty]
# Original snippet
def print_runs(query): 
    if query is None:
        return
    for tup in query:
        print(("{0} @ {1} - {2} id: {3} group: {4}".format(
            tup.end, tup.experiment_name, tup.project_name, 
            tup.experiment_group, tup.run_group)))
\end{lstlisting}

\begin{lstlisting}[style=pythonpretty]
# Modified snippet
def off_runs(number): 
    if number is None:
        return
    for to in number:
        print(("{0} @ {1} - {2} id: {3} group: {4}".format(
            to.end, to.experiment_name, to.project_name, 
            to.experiment_group, to.run_group)))
\end{lstlisting}

\textbf{Identifier Consistency.} This constraint ensures that all occurrences of the same variable or function name are replaced consistently, even when tokenizer artifacts attach different characters to the identifier. If $C_{t_i}$ and $C_{t_j}$ are tokens from different occurrences of the same identifier, let $L_{ij}=\LCS(C_{t_i}, C_{t_j})$ and $L'_{ij}=\LCS(C'_{t_i}, C'_{t_j})$. The constraint is then:
\begin{align}
\Remove(C'_{t_i}, L'_{ij}) &= \Remove(C_{t_i}, L_{ij}), \nonumber \\
\Remove(C'_{t_j}, L'_{ij}) &= \Remove(C_{t_j}, L_{ij}). \nonumber
\end{align}

For instance, the OASIS tokenizer includes different surrounding characters in the tokens corresponding to the two occurrences of the identifier \tokenbox{query}. The original and modified tokens are:
\begin{itemize}
    \item On line 2, let $C_{t_i}=$ \tokenbox{(query}. After replacing \tokenbox{query} with \tokenbox{number}, the corresponding token becomes $C'_{t_i}=$ \tokenbox{(number}.
    \item On line 5, let $C_{t_j}=$ \tokenbox{\textvisiblespace query}, where \tokenbox{\textvisiblespace} denotes the leading whitespace. After replacement, the token becomes $C'_{t_j}=$ \tokenbox{\textvisiblespace number}.
\end{itemize}

Although the complete tokens differ because of their surrounding characters, their longest common substrings isolate the identifier itself: $L_{ij}=$ \tokenbox{query} before the attack and $L'_{ij}=$ \tokenbox{number} afterward. Removing these identifier substrings leaves the surrounding token context unchanged:
\begin{itemize}
    \item For line 2, $\Remove(C'_{t_i}, L'_{ij})=\Remove(C_{t_i}, L_{ij})=$ \tokenbox{(}.
    \item For line 5, $\Remove(C'_{t_j}, L'_{ij})=\Remove(C_{t_j}, L_{ij})=$ \tokenbox{\textvisiblespace}.
\end{itemize}
Thus, the constraint permits the identifier replacement while preserving the parenthesis and whitespace attached by the tokenizer.

\textbf{No Duplicate Replacement.} This constraint guarantees that distinct variables or functions are replaced by distinct new identifiers. If $C_{t_i},C_{t_j}$ are tokens from occurrences of one identifier, and $C_{t_p}, C_{t_q}$ are tokens from occurrences of a \emph{different} identifier, then:
$$\LCS(C'_{t_i}, C'_{t_j}) \neq \LCS(C'_{t_p}, C'_{t_q}).$$

In the example, the tokens are divided into two groups according to the original identifier from which they are derived:
\begin{itemize}
    \item For the original identifier \tokenbox{query}, let $C_{t_i}=$ \tokenbox{(query} and $C'_{t_i}=$ \tokenbox{(number} on line 2. Similarly, let $C_{t_j}=$ \tokenbox{\textvisiblespace query} and $C'_{t_j}=$ \tokenbox{\textvisiblespace number} on line 5.
    \item For the distinct original identifier \tokenbox{tup}, let $C_{t_p}=$ \tokenbox{\textvisiblespace tup} and $C'_{t_p}=$ \tokenbox{\textvisiblespace to} on line 5. The same tokens occur again on line 7, where $C_{t_q}=$ \tokenbox{\textvisiblespace tup} and $C'_{t_q}=$ \tokenbox{\textvisiblespace to}.
\end{itemize}

The first pair therefore represents the replacement \tokenbox{query} $\rightarrow$ \tokenbox{number}, whereas the second represents \tokenbox{tup} $\rightarrow$ \tokenbox{to}. To prevent the two distinct original identifiers from being mapped to the same new identifier, the constraint enforces
\[
\LCS(C'_{t_i}, C'_{t_j}) \neq \LCS(C'_{t_p}, C'_{t_q}),
\]
which in this example reduces to \tokenbox{number} $\neq$ \tokenbox{to}.

Together, these constraints ensure the code's AST structure remains identical. To satisfy the constraints, we employ the Hungarian Algorithm~\citep{HungarianAlgorithm} to match original identifier tokens with their optimal replacements.

Let $V$ denote the tokenizer's vocabulary and $S \subseteq V$ be the set of distinct tokens in the identifiers in the code snippet. For each original token $s_i \in S$, we define an influence function $f_{s_i}: V \rightarrow \mathbb{R}$ derived from gradient information. This function, $f_{s_i}(v)$, quantifies the \textit{influence} when $s_i$ is replaced by $v$.

The goal is to find a set of matching $\{s_i, v_i\}$ that maximizes the total influence, subject to the constraint that each substitute token $v_i$ must be unique. This can be formulated as the following optimization problem:
\begin{equation}
\begin{aligned}
\underset{\{v_i\}}{\text{maximize}} \quad
& \sum_i f_{s_i}(v_i), \\
\text{subject to} \quad
& v_i \in V, \\
& v_i \neq v_j \quad \forall i \neq j.
\end{aligned}
\end{equation}

\subsection{Search Space}
\label{appendix: search_space}

For each token position, we may need to consider up to $|\mathcal{V}|$ candidate tokens to select the replacement token that maximizes the objective in Eq.~\ref{eq: token_replacement_final} (i.e., the combined influence and query-similarity score defined in the methodology). Here, we follow the notation in Section~\ref{ssec: gradient_based_method}: $\mathcal{V}$ denotes the model's full tokenizer vocabulary, and $V \subseteq \mathcal{V}$ denotes the subset of identifier-compatible tokens that satisfies the style preservation requirement used for replacement. In the worst case, $|V|$ can be as large as $|\mathcal{V}|$, but in practice identifier constraints reduce the search space substantially; typically, only about 30--40\% of the full vocabulary is considered.

\subsection{Specification}
The first step of our adversarial attack is to parse the code snippet and locate identifiers suitable for modification. We use language-specific AST parsers: Python's built-in \texttt{ast} library, \texttt{clang} for C++, and \texttt{tree-sitter} for Java, JavaScript, and Go. After applying adversarial modifications, we re-run the parsers to ensure the modified code snippet remains syntactically valid.

For the surrogate models, a replacement token is considered valid only if it consists of characters permitted in standard identifiers across programming languages. However, OASIS presents a unique challenge due to its significantly larger vocabulary: tokens representing an identifier may include preceding operators. For example, a variable \texttt{x} might be tokenized as ``\texttt{\textvisiblespace x}'' (a \textbf{single} token containing the leading white space) and ``\texttt{+x}'' (another \textbf{single} token containing the leading plus sign) at two occurrences in the code snippet. When searching for a replacement variable \texttt{y}, we strictly filter for candidate tokens that preserve this structure---``\texttt{\textvisiblespace y}'' and ``\texttt{+y}'', respectively---and select the one that maximizes influence. Consequently, the search space for OASIS extends to tokens containing operators followed by alphanumeric characters.

% \clearpage
\section{Supplementary Experimental Results}
% \label{appendix: experimental_details_and_supplementary}

\subsection{``Embedding Density'' across Models}
\label{appendix:density_of_embedding}

\begin{table*}[t]
\centering
\caption{Average similarity scores and query--code separation gaps for different models on CosQA and CLARC.}
\label{tab:similarity_density_and_gap}
\small
\setlength{\tabcolsep}{10pt}
\begin{tabular}{ l l c c c } % l=left, c=center; @{} removes extra padding
\toprule
Dataset & Model & \makecell{Avg. Sim between Query \\ and GroundTruth Code} & \makecell{Avg. Sim between Query \\ and Irrelevant Code} & Gap \\
\midrule
\multirow{4}{*}{CosQA} % Dataset spans 4 model rows
 & CodeT5+         & 54.15 & 20.60  & 33.55 \\
 & OASIS           & 68.53 & 47.13 & 21.40  \\
 & Nomic-embed-code  & 41.77 & 1.95  & 39.82 \\
 & Voyage-code-3   & 71.43 & 37.83 & 33.60  \\
\midrule % Rule separating the datasets
\multirow{4}{*}{CLARC} % Dataset spans 4 model rows
 & CodeT5+         & 52.91 & 25.80  & 27.11 \\
 & OASIS           & 85.80  & 54.61 & 31.19 \\
 & Nomic-embed-code  & 55.71 & 7.44  & 48.27 \\
 & Voyage-code-3   & 73.93 & 40.08 & 33.85 \\
\bottomrule
\end{tabular}
\end{table*}

In this subsection, we use the term \emph{embedding density} informally to describe how concentrated query--code similarity scores are in the embedding space. We quantify this using the \emph{gap} between the average similarity for query--ground-truth pairs and for query--irrelevant pairs: a smaller gap indicates a ``denser'' similarity distribution (i.e., weaker separation between relevant and irrelevant snippets).

Table~\ref{tab:similarity_density_and_gap} reports these averages and gaps. Overall, CosQA tends to exhibit a more ``dense'' similarity distribution for some models, largely because irrelevant snippets can still attain relatively high similarity to the query. This is most visible for \texttt{OASIS}, where the gap is comparatively small (e.g., 21.40 on CosQA). Such compressed separation is consistent with our observation in the main paper that attacks evaluated on \texttt{OASIS} tend to yield smaller $\Delta$ Sim than on \texttt{CodeT5+}.

In contrast, CLARC generally shows larger gaps (e.g., 31.19 for \texttt{OASIS}), suggesting a larger ``semantic safety margin'' between ground-truth and irrelevant snippets. This aligns with the weaker overall attack impact on CLARC reported in the paper: adversarially modified irrelevant code must bridge a larger baseline separation to displace the ground truth.

\subsection{Details about Metrics}
\label{appendix: details_about_metrics}
\paragraph{Correlation Metrics}
\begin{itemize}
    \item \textbf{Precision}: The ratio that an attack induced a positive $\Delta$ Similarity on the Surrogate Model also successfully increases similarity on the Eval Model.
    \item \textbf{Pearson Correlation Coefficient ($r$)}: Measures the linear correlation between the numerical values of the similarity changes observed on the Attack and Eval Models.
    \item \textbf{Spearman's Rank Correlation Coefficient ($\rho$)}: Measures the monotonic correlation, assessing how well the rank order of similarity changes is preserved between the Attack and Eval Models.
\end{itemize}

% \subsection{Details about Retrieval Metrics}
\paragraph{Retrieval Metrics}
\begin{itemize}
    \item \textbf{Recall@k (R@k)}: The proportion of queries for which the correct code snippet is found within the top-k ranked results returned by the model. Since each query in the CosQA and CLARC datasets has exactly one ground-truth matching code snippet, R@k specifically measures the percentage of queries where this single correct snippet appears among the top k candidates.

    \item \textbf{Normalized Discounted Cumulative Gain (NDCG)}: A metric for evaluating the quality of a ranked list. It assigns higher scores when relevant items are placed higher in the ranking, applying a logarithmic discount based on position. The score is normalized against the ideal ranking, resulting in a value between 0 and 1.

    \item \textbf{Mean Reciprocal Rank (MRR)}: The average of the reciprocal ranks across all queries in the test set. For a single query, the reciprocal rank is the inverse of the rank position (1/rank) of the ground truth code snippet.
\end{itemize}

\subsection{Distribution of Code Similarity Changes}
\label{appendix: distribution_of_code_similarity_changes}

\begin{figure}[t]
    \centering
    
    % --- Left Subplot ---
    \begin{subfigure}{0.45\textwidth} % Width set to 45% of text width
        \centering
        % \Description{Histogram of cosine similarity changes on CosQA after the attack, separated by evaluation model; positive changes dominate.}
        % Image width set to fill the subfigure container
        \includegraphics[width=\linewidth]{plots/cosqa_codet5p_attack_updated.png}
        \caption{CosQA}
        \label{fig:cosqa_attacked_by_codet5p_in_appendix}
    \end{subfigure}
    \hfill % Adds flexible space between the two images
    % --- Right Subplot ---
    \begin{subfigure}{0.45\textwidth} % Width set to 45% of text width
        \centering
        % \Description{Histogram of cosine similarity changes on CLARC after the attack, separated by evaluation model; most shifts are positive with few negative outliers.}
        % Image width set to fill the subfigure container
        \includegraphics[width=\linewidth]{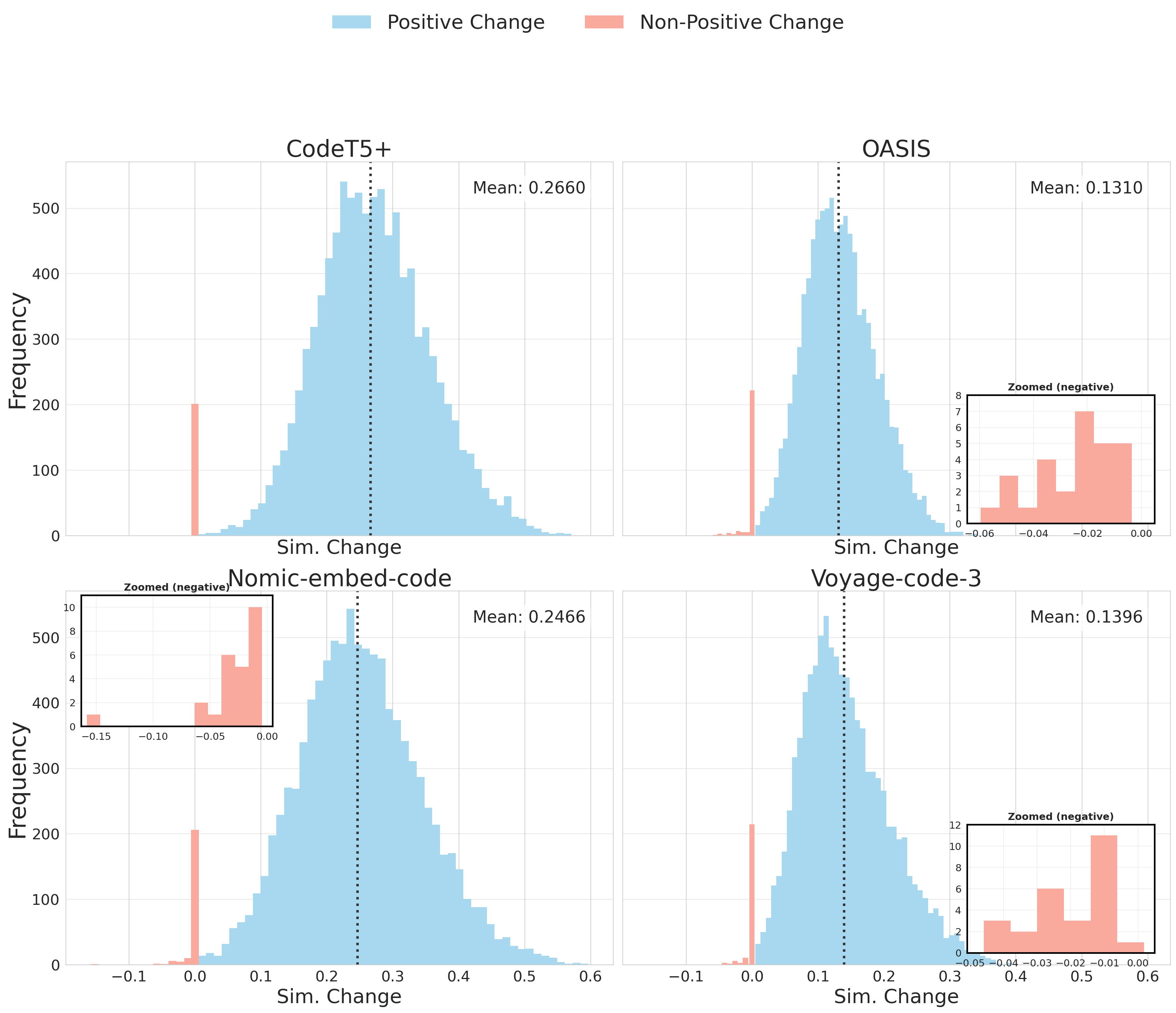}
        \caption{CLARC}
        \label{fig:clarc_attacked_by_codet5p_in_appendix}
    \end{subfigure}

    \caption{Distribution of code similarity changes. The surrogate model is \texttt{CodeT5+}, and the eval models are labeled at the top of each subplot.}
    \label{fig:code_sim_change_in_appendix}
\end{figure}

Figure~\ref{fig:code_sim_change_in_appendix} illustrates the distribution of cosine similarity changes ($\Delta \text{Sim}$) induced by the adversarial code snippets on the CosQA and CLARC datasets. We visualize the results across four distinct embedding models: CodeT5+, OASIS, Nomic-embed-code, and Voyage-code-3.

\textbf{Prevalence of Positive Shifts.} 
The histograms demonstrate that the vast majority of adversarial perturbations result in a positive increase in similarity scores (shown in blue). As indicated by the zoomed-in parts, instances of non-positive change (shown in red), where the adversarial code failed to increase similarity, are almost negligible. The trend confirms the attack's high transferability and consistency across different model architectures.

\textbf{Dataset Comparison.} 
Comparing the two datasets, we observe that the magnitude of $\Delta$ Sim. is consistently larger in CosQA (Python) than in CLARC (C++). For example, the mean $\Delta \text{Sim}$ on CosQA ranges from approximately $0.18$ to $0.39$, whereas on CLARC, the means are more constrained, ranging from $0.13$ to $0.27$. This suggests that the generated adversarial perturbations are generally more effective at inducing large shifts in the embedding space for Python code snippets compared to C++.

\textbf{Model Sensitivity.}
There is also a notable variance in model sensitivity. \texttt{Nomic-embed-code} and \texttt{CodeT5+} consistently exhibit the highest mean similarity shifts. In contrast, \texttt{OASIS} and \texttt{Voyage-code-3} show more conservative shifts. The magnitude of similarity shift can also be explained by \textit{embedding density} from Appendix~\ref{appendix:density_of_embedding}.

\subsection{Query Transferability}
\label{ssec: query_transferability}

\begin{table}[t]
\centering
\caption{$\Delta$ Sim. caused by attacked code snippets when evaluated against original queries (Ori. Q) versus paraphrased queries (Para. Q). The closely matching $\Delta$ Sim. values across both query types demonstrate that our adversarial attack transfers effectively to queries sharing the same semantic intent.}
\label{tab:q_transferability}
\scriptsize
\sisetup{
  separate-uncertainty = true,
  table-align-uncertainty = true
}
\resizebox{\linewidth}{!}{%
\setlength{\tabcolsep}{2pt}
\begin{tabular}{ll S[table-format=2.2(2)] S[table-format=2.2(2)]}
\toprule
Dataset & Eval Model & {$\Delta$ Sim. Ori. Q} & {$\Delta$ Sim. Para. Q} \\
\midrule
\multirow{4}{*}{\textbf{CosQA}}
& CodeT5+       & 38.86 \pm 10.44 & 33.33 \pm 12.03 \\
& OASIS         & 18.15 \pm 6.03  & 17.14 \pm 6.22  \\
& Nomic         & 39.25 \pm 10.89 & 34.96 \pm 12.04 \\
& Voyage-code-3 & 27.13 \pm 8.96  & 25.23 \pm 9.24  \\
\midrule
\multirow{4}{*}{\textbf{CLARC}}
& CodeT5+       & 11.20 \pm 8.66  & 10.58 \pm 8.81  \\
& OASIS         & 13.54 \pm 5.46  & 13.00 \pm 5.38  \\
& Nomic         & 17.78 \pm 8.36  & 17.35 \pm 8.27  \\
& Voyage-code-3 & 11.81 \pm 6.42  & 12.49 \pm 6.69  \\
\bottomrule
\end{tabular}%
}
\end{table}

In practice, developers rarely use identical phrasing for the same semantic intent, meaning a robust attack must generalize across lexical variations. To evaluate this, we use GPT-5.4-mini to generate three paraphrases for each original query in our dataset. We manually validate that the exact intent is preserved. The paraphrased queries introduce a substantial ratio of new tokens not present in the originals: 41.39\% for CosQA and 34.81\% for CLARC. Crucially, the attacked code snippets are optimized against the original queries. We then compare the $\Delta$ Sim. with respect to the original queries and the $\Delta$ Sim. with respect to the corresponding paraphrases, which the optimization never observes.

As the results in Table~\ref{tab:q_transferability} indicate, the attack remains effective on the paraphrased queries. The $\Delta$ Sim. for paraphrased queries closely follows the $\Delta$ Sim. for the original queries across all evaluated models, demonstrating that our attack is targeting the underlying intent rather than fitting to specific lexical triggers. The experiment implies that an adversary does not need prior knowledge of a victim's exact search query. As long as the victim's query aligns with the target semantic intent, the adversarial code snippet will be successfully retrieved, elevating the real-world threat profile of these attacks.

\begin{table}[t]
\centering
\caption{Cross-model transferability of baseline adversarial attacks. The table details the similarity shifts ($\Delta$ Sim.) and Spearman correlation coefficients ($\rho$) when adversarial snippets generated on a source attack model are evaluated across different target models.}
\label{tab:baseline_transferability}
\small
\resizebox{\linewidth}{!}{
\setlength{\tabcolsep}{2pt}
\begin{tabular}{lllcc}
\toprule
Method & Attack Model & Eval Model & $\Delta$ Sim. & $\rho$ \\
\midrule
\multirow{4}{*}{\textbf{DAMP}}       & \multirow{4}{*}{OASIS}  & OASIS         & 10.5 $\pm$ 5.5  & --   \\
                                     &                         & CodeT5+       & 10.8 $\pm$ 11.8 & 63.8 \\
                                     &                         & Nomic         & 16.2 $\pm$ 11.0 & 79.6 \\
                                     &                         & Voyage & 12.7 $\pm$ 8.4  & 84.1 \\
\midrule
\multirow{4}{*}{\textbf{CodeAttack}} & \multirow{4}{*}{Voyage} & Voyage        & 5.2 $\pm$ 4.2   & --   \\
                                     &                         & CodeT5+       & 7.1 $\pm$ 4.9   & 48.3 \\
                                     &                         & Nomic         & 8.8 $\pm$ 3.7   & 52.7 \\
                                     &                         & OASIS         & 5.6 $\pm$ 3.6   & 49.4 \\
\bottomrule
\end{tabular}}
\end{table}

\subsection{Baseline Transferability}
\label{appendix: baseline_transferability}

To assess the cross-model transferability of the baseline adversarial attacks (DAMP and CodeAttack), we measured the similarity shifts ($\Delta$ Sim.) induced by their modified code snippets when evaluated on target models. We also report the Spearman correlation coefficients ($\rho$) for these transfers, as summarized in Table~\ref{tab:baseline_transferability}.

The results confirm that the baseline attacks exhibit transferability across different models. Nevertheless, comparing Table~\ref{tab:baseline_transferability} with Table~\ref{tab:effectiveness_and_transfer} shows that our method achieves higher similarity changes and stronger correlation coefficients in transfer scenarios. This comparison indicates that our method is more effective and controllable than the established baselines.

\subsection{Attack on Various Programming Languages}
\label{ssec: programming_languages}

\begin{table*}[t]
\centering
\caption{Effectiveness and correlation of adversarial attacks transferred from \texttt{CodeT5+} across programming languages in HumanEval-X. $\Delta$ Sim. reports the mean similarity change $\pm$ standard deviation.}
\label{tab:multi_lang_eval}
\sisetup{
  separate-uncertainty = true,
  table-align-uncertainty = true
}
\small
\begin{tabular}{@{} l l S[table-format=2.2(2)] S[table-format=5.0] S[table-format=3.0] S[table-format=2.2] S[table-format=2.2] S[table-format=2.2] @{}}
\toprule
Language & Eval Model & {$\Delta$ Sim.} & {\makecell{Improved \\ Counts}} & {\makecell{Unimproved \\ Counts}} & {Precision} & {$r$} & {$\rho$} \\
\midrule

% Python Section
\multirow{3}{*}{\textbf{Python}} 
 & CodeT5+          & 27.70 \pm 9.14  & 26366 & 530 & {--}  & {--}  & {--}  \\
 & OASIS            & 12.41 \pm 5.21  & 26302 & 594 & 99.76 & 61.68 & 56.00 \\
 & Nomic-embed-code & 26.47 \pm 10.08 & 26290 & 606 & 99.71 & 62.61 & 55.48 \\
\midrule

% C++ Section
\multirow{3}{*}{\textbf{C++}} 
 & CodeT5+          & 24.40 \pm 8.57  & 26823 & 73  & {--}  & {--}  & {--}  \\
 & OASIS            & 14.46 \pm 6.04  & 26667 & 229 & 99.42 & 51.44 & 49.62 \\
 & Nomic-embed-code & 26.13 \pm 10.06 & 26650 & 246 & 99.36 & 50.39 & 48.40 \\
\midrule

% Java Section
\multirow{3}{*}{\textbf{Java}} 
 & CodeT5+          & 24.10 \pm 7.38  & 26865 & 31  & {--}  & {--}  & {--}  \\
 & OASIS            & 15.41 \pm 5.88  & 26738 & 158 & 99.53 & 49.60 & 47.71 \\
 & Nomic-embed-code & 28.10 \pm 10.24 & 26711 & 185 & 99.43 & 48.12 & 45.77 \\
\midrule

% JavaScript Section
\multirow{3}{*}{\textbf{JavaScript}} 
 & CodeT5+          & 28.58 \pm 8.14  & 26866 & 30  & {--}  & {--}  & {--}  \\
 & OASIS            & 16.16 \pm 6.36  & 26720 & 176 & 99.46 & 55.42 & 52.89 \\
 & Nomic-embed-code & 30.76 \pm 10.99 & 26718 & 178 & 99.45 & 53.07 & 49.83 \\
\midrule

% Go Section
\multirow{3}{*}{\textbf{Go}} 
 & CodeT5+          & 27.95 \pm 9.01  & 26855 & 41  & {--}  & {--}  & {--}  \\
 & OASIS            & 16.86 \pm 6.39  & 26752 & 144 & 99.62 & 53.97 & 51.52 \\
 & Nomic-embed-code & 29.79 \pm 10.42 & 26733 & 163 & 99.55 & 51.33 & 48.32 \\

\bottomrule
\end{tabular}
\end{table*}

To assess attack transfer across languages, we use the HumanEval-X dataset~\citep{humaneval_x}, which includes 164 queries with solutions in Python, C++, Java, JavaScript, and Go. We create 26,896 \textit{(query, code)} pairs by pairing each query with all non-corresponding solutions. \texttt{CodeT5+} serves as the surrogate model, while \texttt{OASIS} and \texttt{Nomic-embed-code} are used as eval models.

As shown in Table~\ref{tab:multi_lang_eval}, the adversarial attack is effective and transferable across all five languages. On the surrogate model, all languages exhibit consistently high similarity changes and positive change counts. Meanwhile, the attack is particularly transferable against Java, JavaScript, and Go, where both \texttt{OASIS} and \texttt{Nomic-embed-code} show larger similarity increases than against Python and C++.

Despite these variations in attack magnitude, the correlation coefficients reveal a distinct pattern. Generally, the correlations are moderate across the five languages, suggesting that the monotonic relationship between the attack's impact on the surrogate model and its transferred impact on eval models is stable across languages. However, we observe that the correlation coefficients for Python are notably higher than those for the other programming languages. We hypothesize that such predictability stems from Python's dominance in pre-training corpora, which leads most code embedding models to learn more robust and more consistent representations for Python than for other languages.

\subsection{Different Attack Ratio on Benchmark}
\label{appendix: application_on_benchmark_appendix}

\begin{table*}[t]
\centering
\caption{Model performance under different attack percentages on CosQA. Values report absolute metric drops from the unattacked baseline.}
\label{tab:application_on_benchmark_with_diff_percentage}
\small
\begin{tabular}{lccccccc}
\toprule
{Model Name} & \makecell{\% of Corpus\\Attacked} & \makecell{MRR\\Difference} & \makecell{NDCG\\Difference} & \makecell{R@1\\Difference} & \makecell{R@5\\Difference} & \makecell{R@10\\Difference} & \makecell{R@20\\Difference}  \\
\midrule
\multirow{4}{*}{\makecell{CodeT5+\\(Surrogate)}}
 & 1\%  & 53.54 & 42.28 & 60.00 & 54.00 & 3.20  & 1.00 \\
 & 2\%  & 64.99 & 61.27 & 61.60 & 76.60 & 45.80 & 1.80 \\
 & 5\%  & 71.28 & 74.28 & 62.60 & 83.40 & 83.00 & 71.00 \\
 & 10\% & 72.29 & 76.06 & 62.80 & 85.80 & 87.40 & 86.40 \\
\cmidrule(lr){1-8}
\multirow{4}{*}{OASIS} 
 & 1\%  & 44.29 & 34.80 & 55.80 & 21.80 & 3.80  & 0.40 \\
 & 2\%  & 56.98 & 48.33 & 62.40 & 52.60 & 17.80 & 1.00 \\
 & 5\%  & 69.09 & 67.77 & 66.00 & 74.80 & 62.00 & 30.00 \\
 & 10\% & 74.58 & 75.87 & 68.40 & 83.40 & 79.00 & 63.60 \\
\cmidrule(lr){1-8}
\multirow{4}{*}{Nomic} 
 & 1\%  & 51.26 & 40.09 & 62.00 & 36.00 & 2.80  & 0.40 \\
 & 2\%  & 63.35 & 55.67 & 65.80 & 66.80 & 27.60 & 1.20 \\
 & 5\%  & 73.51 & 73.36 & 69.00 & 77.60 & 71.80 & 45.00 \\
 & 10\% & 77.07 & 78.42 & 70.80 & 85.00 & 81.80 & 72.80 \\
\cmidrule(lr){1-8}
\multirow{4}{*}{Voyage-code-3} 
 & 1\%  & 43.91 & 33.79 & 58.80 & 19.40 & 1.20  & 0.40 \\
 & 2\%  & 57.08 & 46.84 & 66.40 & 45.00 & 11.60 & 0.60 \\
 & 5\%  & 71.34 & 66.97 & 72.80 & 70.80 & 51.20 & 21.40 \\
 & 10\% & 77.54 & 76.63 & 75.00 & 82.00 & 72.60 & 53.20 \\
\bottomrule
\end{tabular}
\end{table*}

We investigated the sensitivity of code retrieval models to varying adversarial attack ratios by evaluating attacks of 1\%, 2\%, and 5\% of the corpus, alongside the original 10\% baseline (Table~\ref{tab:application_on_benchmark_with_diff_percentage}). The experiments were conducted on a corpus of 500 snippets. Consequently, the 1\% attack represents a strictly constrained threat model involving the manipulation of only 5 snippets.

\textbf{Efficacy at low attack ratios.}
The results indicate that the attack remains highly potent even under minimal perturbations. Manipulating 1\% of the candidate pool triggers substantial performance degradation in both white-box and black-box settings. For instance, Recall@1 drops by approximately 60\% on the surrogate model (\texttt{CodeT5+}). Notably, this vulnerability transfers to the commercial embedding model \texttt{Voyage-code-3}, which suffers a 62.8\% drop in Recall@1 under the same conditions. This demonstrates that code search benchmarks' robustness can be compromised by corrupting only a small fraction of the corpus.

\textbf{Trend analysis and saturation.}
As the poisoning percentage increases from 1\% to 5\%, the negative impact on retrieval metrics raises, yet it exhibits a saturation effect. The degradation observed at the 5\% attack rate is nearly equivalent to the drops under the 10\% scenario in the main text. The plateau suggests that our method maximizes the similarity of irrelevant snippets so effectively that a small, strategic injection of adversarial examples (5\% or less) is often sufficient to dominate the top-ranked results.

\subsection{Application of the Adversarial Attack on Other Benchmarks}
\label{appendix:application_of_advattack_on_clarc}

\paragraph{Attacking CLARC}

\begin{table*}[t] % Table placement options: here, top, bottom, page
\centering
\caption{Retrieval metric changes on CLARC caused by adversarial attacks.} % Updated caption
\label{tab:clarc_benchmark_attack} % Updated label
\small
\setlength{\tabcolsep}{12pt}

\begin{tabular}{@{} l l cccccc @{}} % Using left-alignment (l) for Model Name column
\toprule
Model Name & Setting & MRR & NDCG & R@1 & R@5 & R@10 & R@20 \\
\midrule

% CodeT5 (code_t5p)
\multirow{3}{*}{\makecell{CodeT5+\\(Surrogate)}}
 & Original     & 58.84 & 64.55 & 47.34 & 74.14 & 82.51 & 89.54 \\
 & Adversarial  & 2.16 & 3.39  & 0.76  & 3.42  & 7.61  & 16.73 \\
 & $\Delta$     & 56.68 & 61.15 & 46.58 & 70.72 & 74.91 & 72.81 \\
\midrule

% OASIS (oasis)
\multirow{3}{*}{OASIS}
 & Original     & 86.54 & 89.08 & 79.85 & 94.11 & 96.77 & 98.48 \\
 & Adversarial  & 65.95 & 71.15 & 54.37 & 81.18 & 87.26 & 92.40 \\
 & $\Delta$     & 20.59 & 17.93 & 25.48 & 12.93 & 9.51  & 6.08 \\
\midrule

% Nomic-embed-code (nomic)
\multirow{3}{*}{Nomic-embed-code}
 & Original     & 86.23 & 88.61 & 80.04 & 94.11 & 95.82 & 96.96 \\
 & Adversarial  & 55.96 & 61.37 & 45.63 & 68.82 & 78.52 & 84.98 \\
 & $\Delta$     & 30.28 & 27.25 & 34.41 & 25.29 & 17.30 & 11.98 \\
\midrule

% Voyage-code-3 (voyage_ai)
\multirow{3}{*}{Voyage-code-3}
 & Original     & 86.92 & 88.98 & 80.99 & 94.30 & 95.06 & 97.53 \\
 & Adversarial  & 61.26 & 67.43 & 48.48 & 79.28 & 86.69 & 91.45 \\
 & $\Delta$     & 25.66 & 21.54 & 32.51 & 15.02 & 8.37  & 6.08 \\
\bottomrule
\end{tabular}
\end{table*}

Besides CosQA, we also evaluated the impact of adversarial attacks on the CLARC dataset (Table~\ref{tab:clarc_benchmark_attack}). While the attack generated by the surrogate model (\texttt{CodeT5+}) still degrades performance, the magnitude of this degradation is notably smaller compared to the CosQA dataset (Table~\ref{tab:application_on_cosqa_summary}). For instance, while \texttt{OASIS} suffers a catastrophic drop of over 68\% in Recall@1 on CosQA, it retains significantly more robustness on CLARC, with a smaller drop of roughly 25\%.

This disparity in attack effectiveness stems primarily from the intrinsic properties of the datasets:

\begin{itemize}
    \item \textbf{Larger Semantic Safety Margin:} As shown in Table~\ref{tab:similarity_density_and_gap}, the initial similarity gap between the query-ground truth pairs and query-irrelevant pairs is generally wider in CLARC for the transfer models. For example, \texttt{OASIS} exhibits a similarity gap of 31.19 on CLARC compared to only 21.40 on CosQA. This larger initial margin creates a higher barrier for the attacker; the adversarial perturbation must bridge a significantly wider semantic distance to displace the ground truth from the top rank.
    
    \item \textbf{Attack transferability limits:} The adversarial examples generated by the surrogate attack are less transferable (in terms of $\Delta$ Sim.) on CLARC, as shown in Table~\ref{tab:effectiveness_and_transfer}. As a result, the adversarial code snippets often fail to reach similarity scores high enough to crowd out the ground truth during reranking.
\end{itemize}

\paragraph{Attacking CodeSearchNet}

To further demonstrate the generalizability of our approach on a large-scale corpus, we evaluated our method on CodeSearchNet~\cite{CodeSearchNet}. Using CodeT5+ as the surrogate model, we randomly sampled 1,000 (query, code) pairs and applied our attack. We measured the similarity improvement ($\Delta$ Sim.), the number of pairs with positive similarity changes, and Spearman's rank correlation ($\rho$) across multiple evaluation models.

\begin{table}[t]
\centering
\caption{Attack Effectiveness and Transferability on CodeSearchNet. All attacks are using CodeT5+ as the surrogate model.}
\label{tab:eval_model_results}

\small
\setlength{\tabcolsep}{4pt}
\begin{tabular}{lccc}
\toprule
Eval Model & $\Delta$ Sim. & Pos. Change Counts & $\rho$ \\
\midrule
CodeT5+       & $42.5 \pm 10.5$ & 1,000 & --   \\
OASIS         & $23.7 \pm 7.1$  & 1,000 & 54.9 \\
Nomic         & $42.2 \pm 10.9$ & 1,000 & 60.4 \\
Voyage-code-3 & $25.9 \pm 8.1$  & 1,000 & 55.6 \\
\bottomrule
\end{tabular}
\end{table}

The results in Table~\ref{tab:eval_model_results} confirm that our method remains consistently effective and highly transferable on a different corpus. Additionally, the correlation coefficients align closely with those reported in Table~\ref{tab:effectiveness_and_transfer}, indicating that the attack's transferability remains stable and predictable across different datasets.

\subsection{Robust Finetuning}
\label{appendix: robustness_ft}
\begin{table}[t]
    \centering
    \small
    \caption{Performance and robustness metrics of \texttt{CodeT5+} checkpoints. Finetuning with mixed data preserves retrieval baselines and mitigates static attacks. Atk-P: attack on pretrained \texttt{CodeT5+}; Atk-O: attack transferred from \texttt{OASIS}.}
    \label{tab:mitigation_metrics}
    \resizebox{\linewidth}{!}{%
    \begin{tabular}{l ccc |cc}
        \toprule
        \multirow{2}{*}{Checkpoint} & \multicolumn{3}{c}{Retrieval} & \multicolumn{2}{c}{Robustness ($\Delta$ Sim.)} \\
        \cmidrule(lr){2-4} \cmidrule(lr){5-6}
        & MRR & NDCG & R@5 & Atk-P & Atk-O \\
        \midrule
        Pretrain Baseline & 72.4 & 77.3 & 88.0 & 39.90 & 22.25 \\
        Robust-Only FT    & 34.1 & 37.6 & 42.6 & -22.70 & -15.70 \\
        Mixed FT          & 72.3 & 77.6 & 89.4 & 4.00   & 4.42 \\
        \bottomrule
    \end{tabular}}
\end{table}

\begin{table}[t]
\caption{Effectiveness and transferability when the attack is applied directly to finetuned models.}
\label{tab:ft_direct_attack}
\centering
\small
\resizebox{\linewidth}{!}{%
\begin{tabular}{lcccc}
\toprule
 & \multicolumn{2}{c}{Robust-Only FT} & \multicolumn{2}{c}{Mixed FT} \\
\cmidrule(lr){2-3} \cmidrule(lr){4-5}
Eval Model & $\Delta$ Sim. & $\rho$ & $\Delta$ Sim. & $\rho$ \\
\midrule
CodeT5+ (Finetuned) & 7.14 & - & 34.70 & - \\
OASIS         & 3.87 & 65.78 & 15.54 & 71.30 \\
Nomic         & 6.57 & 69.06 & 29.93 & 76.50 \\
Voyage        & 5.40 & 70.55 & 22.61 & 73.66 \\
\bottomrule
\end{tabular}}
\end{table}

To explore potential defense mechanisms, we investigate whether adversarial finetuning can mitigate the proposed attack. Utilizing the Cross Entropy loss, we construct positive pairs from queries and their corresponding ground-truth code snippets, and employ adversarially modified snippets targeting the queries as hard negative samples for our Robust-Only finetuning (Robust-Only FT) model. To ensure the model maintains baseline retrieval capabilities, we introduce a second variant, Mixed FT, which additionally integrates batches of CosQA training data during the finetuning process.

We first evaluate the retrieval performance on CosQA and the robustness ($\Delta$ Sim.) on a test set disjoint from the training data. Here, we measure robustness against two static attacks: one crafted using the original pretrained \texttt{CodeT5+} model (Atk-P), and the other transferred from \texttt{OASIS} (Atk-O). 

As reported in Table~\ref{tab:mitigation_metrics}, Robust-Only FT exhibits strong resistance to these attacks; both Atk-P and Atk-O yield a negative $\Delta$ Sim., indicating these attacks fail to fool the Robust-Only FT checkpoint. However, the retrieval performance of this checkpoint degrades sharply, with all three retrieval metrics dropping to less than half of those achieved by the pretrained baseline. Conversely, the Mixed FT checkpoint strikes a better balance, maintaining high retrieval performance while demonstrating improved resistance to static attacks compared to the baseline.

We further evaluate white-box attacks directly on both finetuned checkpoints. In this setting, the gradients and query-token similarities are computed using the weights of the finetuned models. As reported in Table~\ref{tab:ft_direct_attack}, the Robust-Only FT checkpoint consistently demonstrates strong robustness against white-box attacks; although the $\Delta$ Sim. is positive, the magnitude of the change remains small. In contrast, the Mixed FT checkpoint proves vulnerable to white-box attacks, yielding $\Delta$ Sim. values comparable to those in Table~\ref{tab:effectiveness_and_transfer}. Notably, the absolute Spearman correlation coefficients ($\rho$) for both checkpoints are larger than those observed in Table~\ref{tab:effectiveness_and_transfer}, suggesting that attack transferability from the finetuned \texttt{CodeT5+} models is more predictable.

Overall, our robust finetuning experiments demonstrate that standard adversarial finetuning is an insufficient mitigation strategy for our attack. The resulting checkpoints suffer from a trade-off: they either remain vulnerable to white-box attacks or achieve robustness at the cost of severely degraded retrieval performance.

\subsection{Ablation Studies}
\label{appendix: ablation_studies} 

\begin{figure*}[t] 
    \centering % Center the image within the figure environment
    % \Description{Line plot of average query--code similarity across attack iterations on CosQA, comparing selecting the best-so-far adversarial snippet versus selecting a fixed-iteration output, evaluated on the surrogate and transfer models.}
    \includegraphics[width=\textwidth]{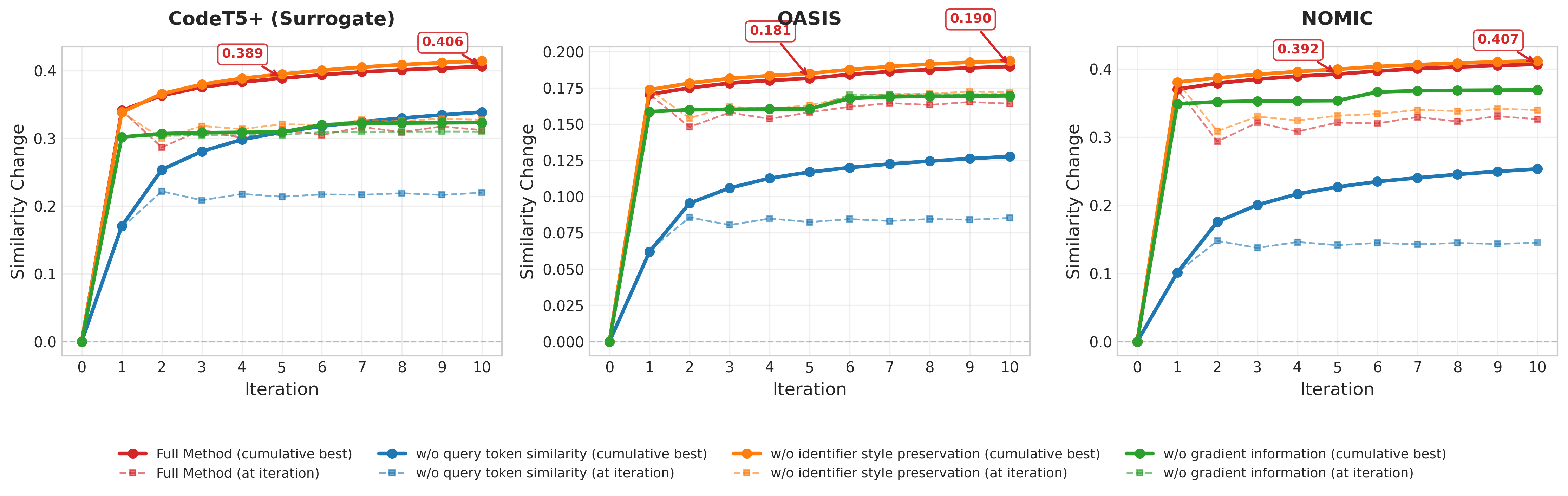}
    \caption{Average similarity change on CosQA using \texttt{CodeT5+} as the surrogate model. We evaluate the attacks on the surrogate itself (left) and two transfer targets: \texttt{OASIS} and \texttt{Nomic-embed-code} (center/right). Selecting the code with maximum similarity up to iteration $k$ leads to higher similarity changes, while extending the attack from 5 to 10 iterations yields only marginal gains in $\Delta$ Sim.} % Caption for the figure
    \label{fig:ablation_sim_change} % Label for cross-referencing
\end{figure*}

We conduct additional ablation studies to evaluate the contribution of specific components within our attack method.

\paragraph{Attack Code Selection}
Our default strategy selects the adversarial code that achieves the highest similarity to the query across all iterations of the attack process. In this ablation, we compare this approach with an alternative that selects the adversarial code directly from a fixed iteration $k$ in order to justify the necessity of our adversarial code selection strategy.

We evaluated both strategies using the 10,000 sampled \textit{(query, code)} pairs from CosQA for 10 iterations. Figure~\ref{fig:ablation_sim_change} plots the average query-code similarity at each iteration $k$ for both selection strategies under the full method and the ablation methods. The alternative approach, which selects the code at iteration $k$ (dashed lines), leads to less effective attacks, as the average similarity plateaus and fluctuates after three iterations. In contrast, selecting the code with the maximum similarity observed up to iteration $k$ (solid lines) allows the average similarity to increase monotonically throughout the process, achieving much higher final similarities. The comparison confirms the benefit of retaining the highest-scoring code variant across all iterations rather than only considering the final iteration's output.

\paragraph{Iteration Limit}
Our default adversarial attack protocol employs 5 iterations. As shown in Figure~\ref{fig:ablation_sim_change} (solid lines), the average query-code similarity continues to increase only marginally beyond this point across all evaluation models. Specifically, extending the process from 5 to 10 iterations yields a $\Delta$ Similarity increase of less than 0.02 in all cases (both surrogate and transfer settings). Given that these additional iterations double the computational cost for diminishing returns, we terminate the attack at 5 iterations. Conversely, for scenarios with strictly limited resources, the steep initial rise in the plots suggests that running the attack for just a single iteration still yields substantial similarity improvements. However, where efficiency is not a primary constraint, performing additional iterations remains beneficial for maximizing attack performance.

\begin{table}[t]
    \caption{Ablation on $\alpha$. All values are $\Delta$ Similarity on 1,000 sampled \textit{(query, code)} pairs from CosQA, scaled by 100.}
    \label{tab: ablation_on_alpha}
    \centering
    \small
    \begin{tabular}{lccc}
        \toprule
        {Model} & {CodeT5+} & {OASIS} & {Nomic-embed-code} \\
        \midrule
        $\alpha=0.01$ & 33.95 & 14.74 & 28.78 \\
        $\alpha=0.05$ & 39.30 & 18.89 & 40.06 \\
        $\alpha=0.1$  & 39.83 & 19.45 & 42.39 \\
        $\alpha=0.5$  & 38.15 & 19.18 & 42.19 \\
        \bottomrule
    \end{tabular}
\end{table}

\paragraph{Choice of $\alpha$} We conduct an additional ablation study to determine the optimal hyperparameter $\alpha$ in Equation~\ref{eq: token_replacement_final}. In this experiment, we sample 1,000 \textit{(query, code)} pairs from CosQA and use \texttt{CodeT5+} as the surrogate model to test $\alpha \in \{0.01, 0.05, 0.1, 0.5\}$. We then evaluate the resulting $\Delta$ Similarity on \texttt{CodeT5+}, \texttt{OASIS}, and \texttt{Nomic-embed-code}.

As reported in Table~\ref{tab: ablation_on_alpha}, \texttt{CodeT5+} exhibits comparable $\Delta$ Similarity scores when $\alpha$=0.05 and $\alpha$=0.1. However, regarding transferability, we observe that \texttt{OASIS} and \texttt{Nomic-embed-code} achieve higher scores with larger $\alpha$ values. Consequently, we selected $\alpha$=0.1 for our main experiments, as it consistently yielded the highest $\Delta$ Similarity across models.

% \clearpage
\section{Adversarial Attack Examples}
\label{appendix: adversarial_attack_examples}
In this section, we provide the code snippets and their target queries before and after the adversarial attack.
% For more examples, please refer to our GitHub repository at \url{https://github.com/AdvAttackOnNCC/Code_Search_Adversarial_Attack}.

\subsection{CosQA}
\label{appendix: CosQA Example}
\textbf{Query Example} \texttt{using sort to move element in to new position in list python
} %\textcolor{white}{Query ID: q20513}

\paragraph{Original Code Snippet} 

\begin{itemize}
    \item Similarity with Query based on CodeT5+: 0.2814
    \item Similarity with Query based on OASIS: 0.5312
    \item Similarity with Query based on Nomic-embed-code: 0.0689
\end{itemize}

%\textcolor{white}{Code ID: d20105}

\begin{lstlisting}[style=pythonpretty]
def process_and_sort(s, force_ascii, full_process=True):
    ts = utils.full_process(s, force_ascii=force_ascii) if full_process else s
    tokens = ts.split()
    sorted_string = u" ".join(sorted(tokens))
    return sorted_string.strip()
\end{lstlisting}

\paragraph{Attacked Code Snippet} 

\begin{itemize}
    \item Similarity with Query based on CodeT5+: 0.7300 \textcolor{Emerald}{($\uparrow$ 0.4486)}
    \item Similarity with Query based on OASIS: 0.7426 \textcolor{Emerald}{($\uparrow$ 0.2114)}
    \item Similarity with Query based on Nomic-embed-code: 0.5925 \textcolor{Emerald}{($\uparrow$ 0.5236)}
\end{itemize}

\begin{lstlisting}[style=pythonpretty]
def newusing_sort_move( element, force_ascii, full_process=True):
    ts = utils.full_process( element, force_ascii=force_ascii) if full_process else element
    move = ts.split()
    position_string = u" ".join(sorted( move))
    return position_string.strip()
\end{lstlisting}

\paragraph{Attacked Code Snippet (w/o gradient)} 

\begin{itemize}
    \item Similarity with Query based on CodeT5+: 0.6653 \textcolor{Emerald}{($\uparrow$ 0.3839)}
    \item Similarity with Query based on OASIS: 0.7554 \textcolor{Emerald}{($\uparrow$ 0.2242)}
    \item Similarity with Query based on Nomic-embed-code: 0.4992 \textcolor{Emerald}{($\uparrow$ 0.4303)}
\end{itemize}

\begin{lstlisting}[style=pythonpretty]
def sortusing_new_element( python, position_ascii, move_process=True):
    position = utils.full_process( python, force_ascii= position_ascii) if move_process else python
    tokens = position.split()
    move_string = u" ".join(sorted(tokens))
    return move_string.strip()
\end{lstlisting}

\paragraph{Attacked Code Snippet (w/o style preservation)} 

\begin{itemize}
    \item Similarity with Query based on CodeT5+: 0.7613 \textcolor{Emerald}{($\uparrow$ 0.4799)}
    \item Similarity with Query based on OASIS: 0.7345 \textcolor{Emerald}{($\uparrow$ 0.2033)}
    \item Similarity with Query based on Nomic-embed-code: 0.5225 \textcolor{Emerald}{($\uparrow$ 0.4536)}
\end{itemize}

\begin{lstlisting}[style=pythonpretty]
def sortelementsortnewusingelementsreviewList( python, force_ascii, new_process=True):
    to = utils.full_process( python, force_ascii=force_ascii) if new_process else python
    element = to.split()
    python_string = u" ".join(sorted( element))
    return python_string.strip()
\end{lstlisting}

\paragraph{Attacked Code Snippet (w/o query token similarity)} 

\begin{itemize}
    \item Similarity with Query based on CodeT5+: 0.5808 \textcolor{Emerald}{($\uparrow$ 0.2994)}
    \item Similarity with Query based on OASIS: 0.6336 \textcolor{Emerald}{($\uparrow$ 0.1024)}
    \item Similarity with Query based on Nomic-embed-code: 0.3208 \textcolor{Emerald}{($\uparrow$ 0.2519)}
\end{itemize}

\begin{lstlisting}[style=pythonpretty]
def oosordering_AttachmentFileEntry_sort( Ordering, getTypes_Random, digest_Processes=True):
    ptr = utils.full_process( Ordering, force_ascii= getTypes_Random) if digest_Processes else Ordering
    duplicates = ptr.split()
    CONT_List = u" ".join(sorted( duplicates))
    return CONT_List.strip()
\end{lstlisting}

\subsection{CLARC}
\label{appendix: CLARC Example}
\textbf{Query Example} \texttt{The function takes a constant string reference as its input and copies its content to another string reference provided as the output. It effectively performs a direct assignment, duplicating the input string without modification. The input is a constant string (std::string) and the output is also a string (std::string).} %\textcolor{white}{Query ID: q_group_1_id_503}

\paragraph{Original Code Snippet} 

\begin{itemize}
    \item Similarity with Query based on CodeT5+:  0.3851
    \item Similarity with Query based on OASIS: 0.6060
    \item Similarity with Query based on Nomic-embed-code: 0.1037
\end{itemize}

%\textcolor{white}{Code ID: c_group_1_id_323}

\begin{lstlisting}[style=cpretty]
void fp_reverse (unsigned char *s, int len)
{
  int     ix, iy;
  unsigned char t;

  ix = 0;
  iy = len - 1;
  while (ix < iy) {
    t     = s[ix];
    s[ix] = s[iy];
    s[iy] = t;
    ++ix;
    --iy;
  }
}
\end{lstlisting}

\paragraph{Attacked Code Snippet} 

\begin{itemize}
    \item Similarity with Query based on CodeT5+: 0.6544  \textcolor{Emerald}{($\uparrow$ 0.2694)}
    \item Similarity with Query based on OASIS: 0.7676 \textcolor{Emerald}{($\uparrow$ 0.1616)}
    \item Similarity with Query based on Nomic-embed-code: 0.5072 \textcolor{Emerald}{($\uparrow$ 0.4036)}
\end{itemize}

\begin{lstlisting}[style=cpretty]
void copies_string (unsigned char * reference, int content)
{
  int     assignment, iy;
  unsigned char output;

  assignment = 0;
  iy = content - 1;
  while ( assignment < iy) {
    output     = reference[ assignment];
    reference[ assignment] = reference[iy];
    reference[iy] = output;
    ++ assignment;
    --iy;
  }
}
\end{lstlisting}

\paragraph{Attacked Code Snippet (w/o gradient)} 

\begin{itemize}
    \item Similarity with Query based on CodeT5+: 0.5427 \textcolor{Emerald}{($\uparrow$ 0.1576)}
    \item Similarity with Query based on OASIS: 0.6856 \textcolor{Emerald}{($\uparrow$ 0.0796)}
    \item Similarity with Query based on Nomic-embed-code: 0.3424 \textcolor{Emerald}{($\uparrow$ 0.2387)}
\end{itemize}

\begin{lstlisting}[style=cpretty]
void output_string (unsigned char * content, int len)
{
  int     assignment, iy;
  unsigned char takes;

  assignment = 0;
  iy = len - 1;
  while ( assignment < iy) {
    takes     = content[ assignment];
    content[ assignment] = content[iy];
    content[iy] = takes;
    ++ assignment;
    --iy;
  }
}
\end{lstlisting}

\paragraph{Attacked Code Snippet (w/o style preservation)} 

\begin{itemize}
    \item Similarity with Query based on CodeT5+: 0.5845 \textcolor{Emerald}{($\uparrow$ 0.1995 )}
    \item Similarity with Query based on OASIS:  0.7493 \textcolor{Emerald}{($\uparrow$ 0.1434)}
    \item Similarity with Query based on Nomic-embed-code: 0.4364 \textcolor{Emerald}{($\uparrow$ 0.3328)}
\end{itemize}

\begin{lstlisting}[style=cpretty]
void modificationConstantstringstd (unsigned char * reference, int len)
{
  int     constant, iy;
  unsigned char output;

  constant = 0;
  iy = len - 1;
  while ( constant < iy) {
    output     = reference[ constant];
    reference[ constant] = reference[iy];
    reference[iy] = output;
    ++ constant;
    --iy;
  }
}
\end{lstlisting}

\paragraph{Attacked Code Snippet (w/o query token similarity)} 

\begin{itemize}
    \item Similarity with Query based on CodeT5+:  0.5256 \textcolor{Emerald}{($\uparrow$ 0.1405)}
    \item Similarity with Query based on OASIS: 0.6942 \textcolor{Emerald}{($\uparrow$ 0.0883)}
    \item Similarity with Query based on Nomic-embed-code: 0.2849 \textcolor{Emerald}{($\uparrow$ 0.1812)}
\end{itemize}

\begin{lstlisting}[style=cpretty]
void Compile_copy (unsigned char * Expressions, int getConstant)
{
  int     begin, iy;
  unsigned char iss;

  begin = 0;
  iy = getConstant - 1;
  while ( begin < iy) {
    iss     = Expressions[ begin];
    Expressions[ begin] = Expressions[iy];
    Expressions[iy] = iss;
    ++ begin;
    --iy;
  }
}
\end{lstlisting}

%%%%%%%%%%%%%%%%%%%%%%%%%%%%%%%%%%%%%%%%%%%%%%%%%%%%%%%%%%%%%%%%%%%%%%%%%%%%%%%
%%%%%%%%%%%%%%%%%%%%%%%%%%%%%%%%%%%%%%%%%%%%%%%%%%%%%%%%%%%%%%%%%%%%%%%%%%%%%%%

\end{document}